\documentclass[prd,aps,reprint,onecolumn,superscriptaddress,tightenlines,nofootinbib,eqsecnum,preprintnumbers,longbibliography,12pt]{revtex4-2}
\usepackage{bbm}
\usepackage{amsmath,amssymb}
\usepackage{physics}
\usepackage{dsfont}
\usepackage{graphicx}
\usepackage{hyperref}
\usepackage[dvipsnames]{xcolor}

\usepackage{bm}

\begin{document}


\title{Influence of new states in searches for negative gauge-Higgs couplings}
\def\TRIUMF{TRIUMF, 4004 Wesbrook Mall, Vancouver, BC V6T 2A3, Canada}
\def\Carleton{Ottawa-Carleton Institute for Physics, Carleton University, Ottawa, ON K1S 5B6, Canada}
\author{Carlos Henrique de Lima}
\email{cdelima@triumf.ca}
\affiliation{\TRIUMF}

\author{Daniel Stolarski}
\email{stolar@physics.carleton.ca}
\affiliation{\Carleton}

\date{\today}
\begin{abstract}
In this work, we explore how constraints based on interference effects for the negative gauge-Higgs coupling scenario are affected by new physics. Models that achieve this wrong-sign gauge-Higgs coupling inevitably have new charged Higgs states. These states affect the interpretation of interference effects. We recast the ATLAS analysis for VBF $Wh$, showing that the previously excluded negative gauge-Higgs coupling scenario is still experimentally viable for charged Higgs masses \textit{below} $370$~GeV. We show that it is possible to weaken this bound further with mild tuning in the parameter space. We discuss how to further constrain the wrong-sign gauge-Higgs coupling hypothesis and point out the potential of VBF $Zh$ to exclude this scenario in a model-independent way.
\end{abstract}


\maketitle

\section{Introduction}

The Standard Model (SM) Higgs unitarizes gauge boson scattering~\cite{Lee:1977yc,Lee:1977eg}. If, however, the Higgs-like particle has couplings that differ from the SM predictions, then new physics that is not much above the electroweak scale is required to make the theory unitary~\cite{Lee:1977yc,Lee:1977eg,Logan:2022uus}. The LHC experiments have made many measurements of the Higgs couplings, all consistent with the SM prediction~\cite{CMS:2022dwd,ATLAS:2022vkf}. This means that the Higgs couplings to the $W$, $Z$, and third-generation fermions must be similar to those in the SM.

Measurements of Higgs couplings typically use rates and are thus proportional to the magnitude of the couplings. Therefore, an intriguing possibility that may not be excluded is when some Higgs couplings have the opposite sign compared to the SM prediction. The sign of the couplings is only probed through interference effects, for example in Higgs decays to four leptons~\cite{Chen:2016ofc}, $W^+W^- h$ production~\cite{Chiang:2018fqf}, VBF $Vh$ production~\cite{Stolarski:2020qim}, and the combination of $Zh$ and $th$ production~\cite{Xie:2021xtl}. The measurements proposed assume that one can vary the sign of the couplings without modifying anything else. While this is a useful first approximation, the unitarity analysis means there must be new states in the theory if the Higgs couplings are modified, especially when involving the gauge sector. Furthermore, since these wrong sign scenarios significantly modify the Higgs couplings, at least one new state must be below about 620 GeV~\cite{Das:2024xre} and thus could significantly change the interference measurements.

Modifications of the gauge-Higgs coupling $\kappa_{V}$ are the most sensitive for new physics as it is required in the unitarization of gauge boson scattering. In this work, we explore the possibility of a negative gauge-Higgs coupling. Specifically, we focus on the ratio of couplings
\begin{align}
\lambda_{WZ} = \frac{\kappa_{W}}{\kappa_{Z}} \, .
\end{align}
In~\cite{Stolarski:2020qim}, it was pointed out that the VBF $Vh$ process, one that is sensitive to the hard process $VV\rightarrow Vh$ with $V=Z, W$, has tree-level interference between diagrams with a coupling of a Higgs to a $W$ (parameterized by $\kappa_W$) and couplings of a Higgs to a $Z$ (parameterized by $\kappa_Z$). Therefore, the rate of this process is very sensitive to the sign of $\lambda_{WZ}$. For longitudinally polarized gauge bosons, the amplitude grows as  $(\lambda_{WZ}-1)s$ where $s$ is the usual Mandelstam variable. These observations motivated experimental analyses by ATLAS~\cite{ATLAS:2024vxc} and CMS~\cite{CMS:2023sdc}, which claim exclusion of $\lambda_{WZ} \approx -1$ with high confidence.

In this work, we explore how the constraints on the $\lambda_{WZ} \approx -1$ scenario are modified by new states. Generating a tree-level (and thus large) modification to $\lambda_{WZ}$ breaks the SM custodial symmetry and can only occur in models with scalar representations larger than doublets~\cite{Veltman:1977kh,Gunion:1989we}. These models contain new charged Higgs states contributing to $VV\rightarrow Vh$ at the tree level. Furthermore, to unitarize $VV$ scattering, the states cannot be arbitrarily heavy. Therefore, one expects significant changes to that process, which can modify the interpretation of experimental results.

Generically, models with scalar representations larger than doublet are heavily constrained by electroweak precision data~\cite{deLima:2022yvn,ParticleDataGroup:2020ssz}. One way to avoid such constraints is to have a vacuum that preserves the custodial symmetry while the custodial symmetry breaking appears only in the interactions and spectrum. This is achieved in the accidentally custodial symmetric (AC) models introduced in~\cite{deLima:2021llm}. The AC models have the same field content as generalized Georgi-Machacek models~\cite{GGM}, with a custodial-symmetry-breaking potential. 

An important point to notice is that there is an extremely vast space of models that can generate a negative gauge-Higgs coupling. Nevertheless, these models all contribute universally to the $VV\rightarrow Vh$ scattering. Therefore, studying this process allows us to experimentally probe \emph{all} models directly. Each individual model may have stronger bounds coming from other processes, but these model-dependent bounds cannot be tested universally. We can parametrize the effect in terms of one additional charged state where the couplings are fixed by unitarity~\cite{Gunion:1990kf}, and we have only the mass and width as free parameters. We show how the new state alleviates the growth of the cross section provided the state is below the energy probed. Generically, the amplitude of $VV\rightarrow Vh$ with longitudinally polarized vectors grows with energy up to an energy scale comparable to the mass of the new states. We can thus reinterpret the bounds on $\lambda_{WZ}$ as a bound for the heaviest mass state that unitarizes the amplitude. The VBF $Vh$ process can then be used to test the hypothesis that one of the gauge-Higgs couplings is negative in a model-independent way. In this parametrization, excluding all charged Higgs masses translates to excluding this hypothesis. 

We recast the ATLAS VBF $Wh$~\cite{ATLAS:2024vxc} analysis that claims exclusion of  $\lambda_{WZ} = -1$ at high confidence. We show that in the presence of new physics states, a wide range of charged masses is still allowed. This channel only starts to differentiate from the SM once the charged Higgs mass exceeds the $Wh$ threshold. Using the ATLAS result, we establish a $95\%$ CL exclusion on masses above $370~\text{GeV}$. This means that it is not possible to entirely exclude the hypothesis of a relative sign between gauge-Higgs couplings with current experimental results. We also note that additional data will only marginally improve this exclusion because the $Wh$ process becomes kinematically similar to the SM for small masses. A complementary model-independent exclusion of some models with negative $\lambda_{WZ}$ was performed in~\cite{deLima:2021llm} using Higgs couplings measurements. Another method to probe the sign of the $hZZ$ coupling was proposed in~\cite{Das:2024xre} using a similar approach but focusing on neutral Higgs searches.

Models with light charged scalars can be constrained by many other experimental searches. The mode most closely related to the process studied here is the classic gauge boson scattering $VV\rightarrow VV$ ($V=W, Z$) with a charged scalar mediating (possibly resonant) modifications to the SM. Other potential constraints are decays of the charged Higgs to SM fermions, and having the charged state participate in loop processes such as $h\rightarrow \gamma\gamma$. We will briefly explore all these processes in this work, but we stress that all the limits will be \textit{model-dependant}. They will depend on different model parameters that do not participate in the unitarization of $VV\rightarrow Vh$. In order to exclude the remaining mass window in a model-independent way, it is necessary to study the VBF $Zh$ process. This channel is kinematically similar to the SM in the low mass window, but the cross section is larger by a factor of at least 1.5 in this regime. VBF $Zh$ has a small cross section, but could potentially be probed with HL-LHC~\cite{Belvedere:2024wzg,Butler:2023eah,ATLAS:2019mfr} or in a future lepton collider~\cite{Bambade:2019fyw,Aicheler:2018arh,Bai:2021rdg}. 

The remainder of this paper is organized as follows. In Sec.~\ref{sec:modelI}, we explore the $2 \rightarrow 2$ process $VV\rightarrow Vh$ ($V=W, Z$) focusing on the effects of new states and describe how these effects can be approximated by one charged Higgs. In Sec.~\ref{sec:ATLAS}, we perform the recast of the ATLAS analysis, and in Sec.~\ref{sec:other}, we describe other model-dependent searches that can constrain the charged Higgs mass. In Sec~\ref{sec:conc}, we summarize our work and discuss tackling the remaining allowed parameter space using the VBF $Zh$ process. In Appendix~\ref{ap}, we discuss how the bounds can be further weakened in tuned scenarios with multiple new states.

\section{Model-independent analysis of $V \, V \rightarrow V \,  h $ } 
\label{sec:modelI}

Here we will explore the effects of new physics in the $VV\rightarrow Vh$ ($V=W, Z$) process. As shown in~\cite{Stolarski:2020qim}, when the Higgs couplings to gauge bosons deviate from their SM values, this process grows with energy and eventually becomes inconsistent with unitarity. This effect is particularly dramatic if all three gauge bosons are longitudinally polarized. Therefore, new physics is required for this theory to be consistent, and this new physics cannot be at arbitrarily high masses. 

In this work, we focus on the wrong sign scenario with 
\begin{equation}
\kappa_W = 1, \; \; \kappa_Z = -1
\label{eq:wrong}
\end{equation}
which implies $\lambda_{WZ}=-1$.\footnote{ The results would be the same if we instead chose $\kappa_W = -1$ and $\kappa_Z = 1$ as the cross section depends only on the magnitude of the couplings and their relative sign.} Previous work~\cite{Stolarski:2020qim} including the experimental results~\cite{ATLAS:2024vxc,CMS:2023sdc} assume that one can take only the SM field content and the modified couplings, but such a scenario is inconsistent with unitarity and one must also add relatively light new states. In particular, weakly coupled models that can give the wrong sign scenario of Eq.~\eqref{eq:wrong} must have a singly charged Higgs scalar. 

We have AC models~\cite{deLima:2021llm} in mind for this analysis, but any consistent weakly coupled extension has similar properties. As noted above, these models have the same field content as those that preserve custodial symmetry. They can, therefore, be analyzed as custodial models deformed to include custodial symmetry breaking. These models have additional custodial triplets and fiveplets\footnote{There can be higher multiplets, but those do not contribute to this process.} in the spectrum. For example, the AC triplet model has the Georgi-Machacek model as its custodial limit with a custodial fiveplet $H_{5}$ and a custodial triplet $H_{3}$. Once we include custodial violation, states from different custodial representations with the same electric charge get mixed. The relevant mixing for the $VV\rightarrow Vh$ process occurs between the singly charged states in $H_{5}$ and $H_{3}$ with couplings to $WZ$ and $Wh$. This mixing can be described by the following custodial symmetry violating Lagrangian for the charged scalars $H^+_i$:
\begin{align}\label{Eq:NPEQ}
\mathcal{L}_{H^{+}} = g \, m_{Z} \, g^{i}_{WZ} H_{i}^{+}W^{-}_{\mu}Z^{\mu} - i \, g \, g^{i}_{Wh}\,  h \, \partial_{\mu}H^{+}_{i} \, W^{- \, \mu}  + \text{h.c.} \, .
\end{align}
Depending on the specific model, we can have a multiplicity $i$ of charged scalars with these custodial-violating interactions. In addition, we also have the standard modification of the Higgs couplings, which we can describe model independently in the $\kappa$-framework
\begin{align}
\mathcal{L}_{\kappa}=h\left(\kappa_W \,  g \, m_W \,  W_\mu^{+} W^{- \, \mu} + \kappa_Z\,  g \frac{m_Z^2}{2 m_W}\,  Z_\mu Z^\mu\right) \, .
\end{align}

\begin{figure}[t!]
 \resizebox{0.2\linewidth}{!}{ \includegraphics{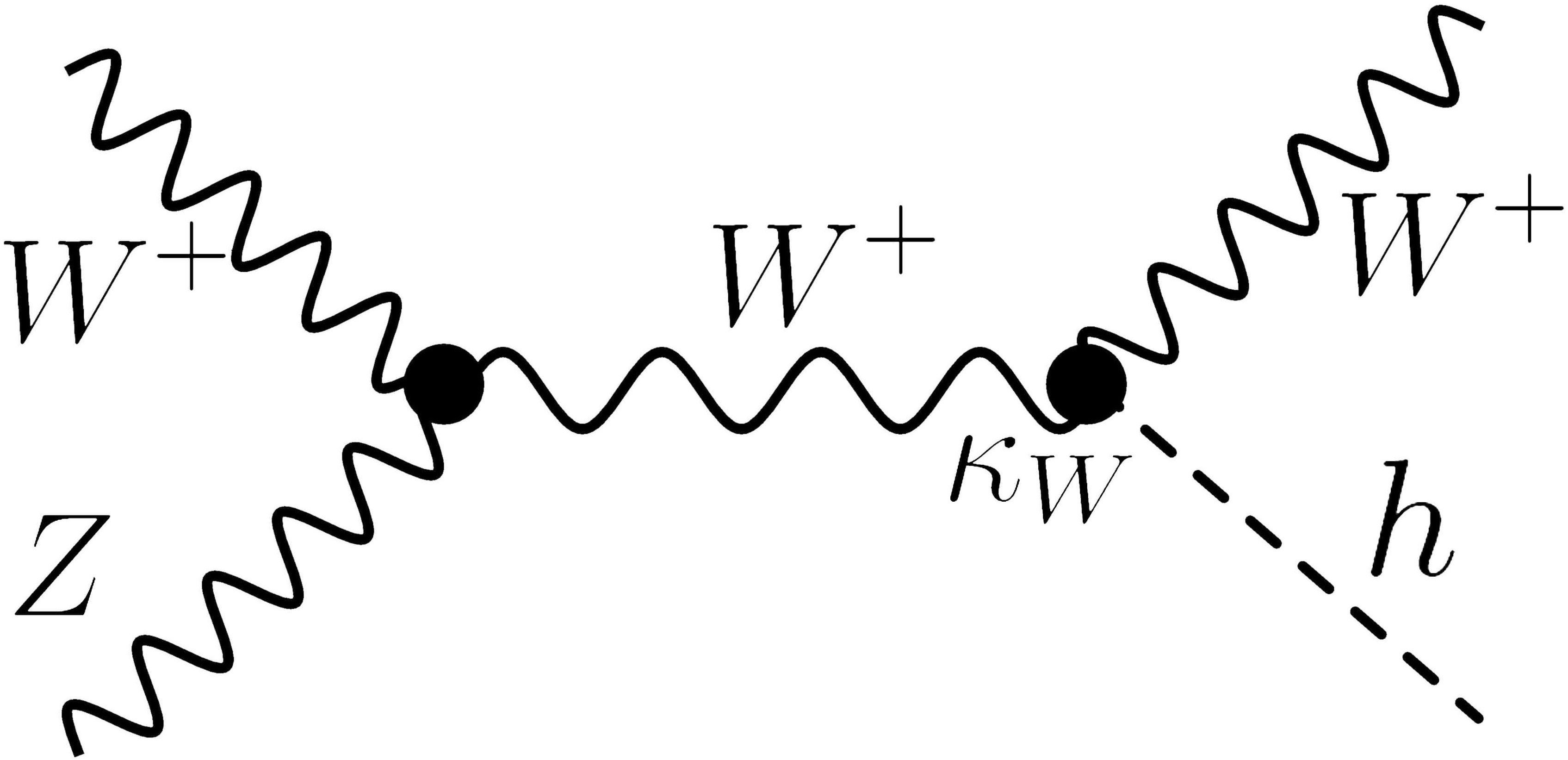}}  \,  
 \resizebox{0.17\linewidth}{!}{ \includegraphics{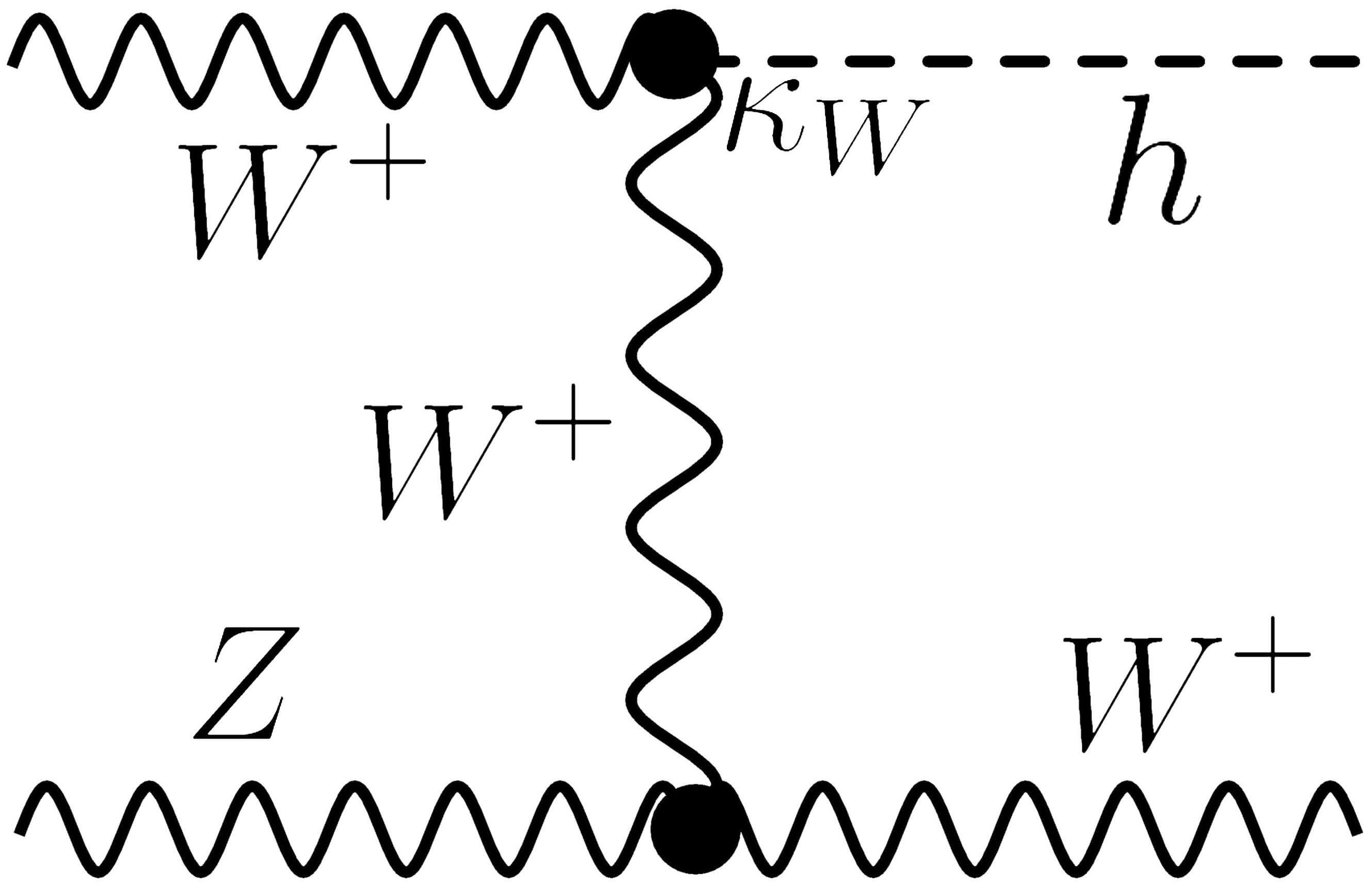}} \,  
 \resizebox{0.17\linewidth}{!}{ \includegraphics{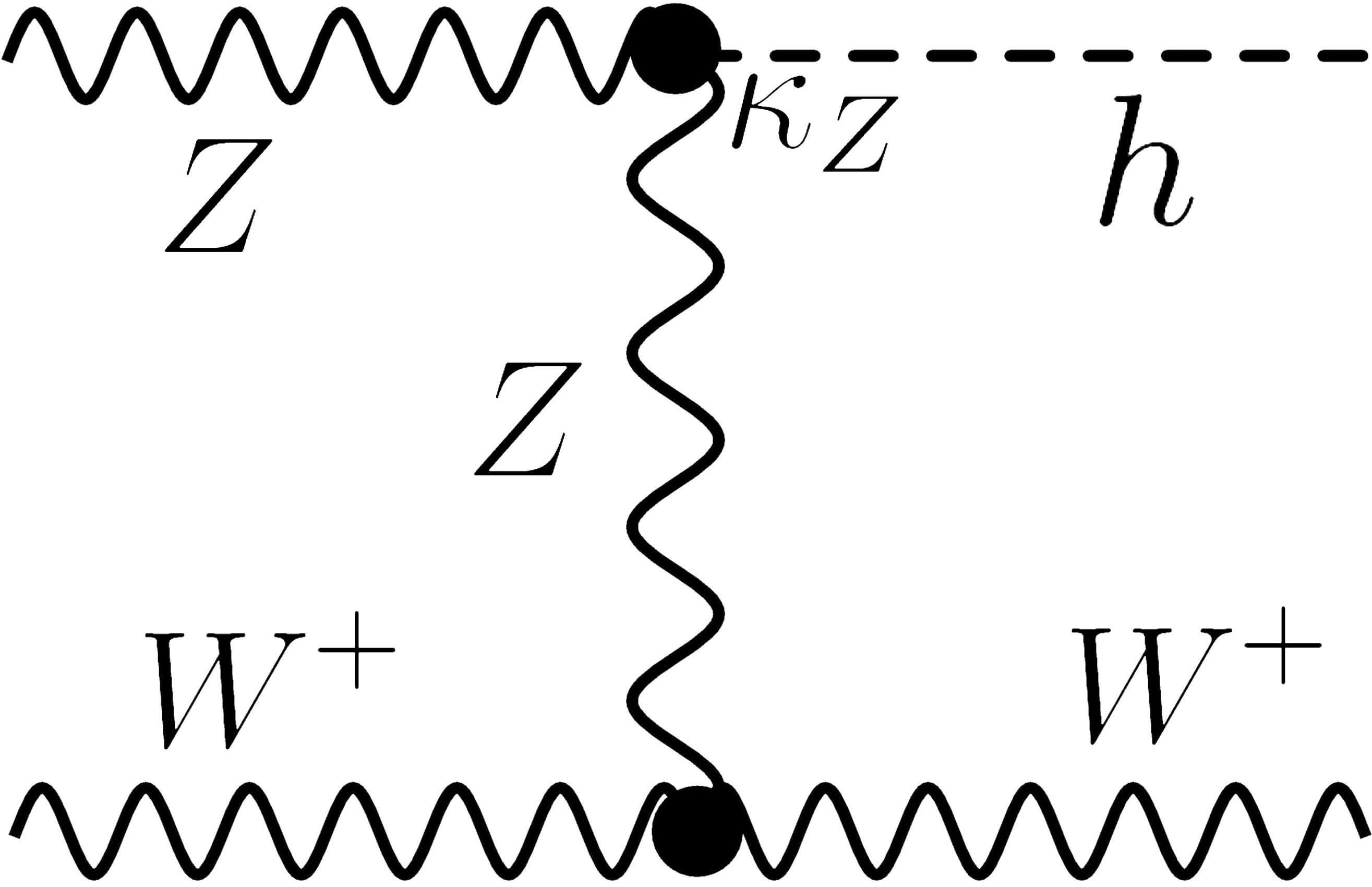}} \,  
 \resizebox{0.17\linewidth}{!}{ \includegraphics{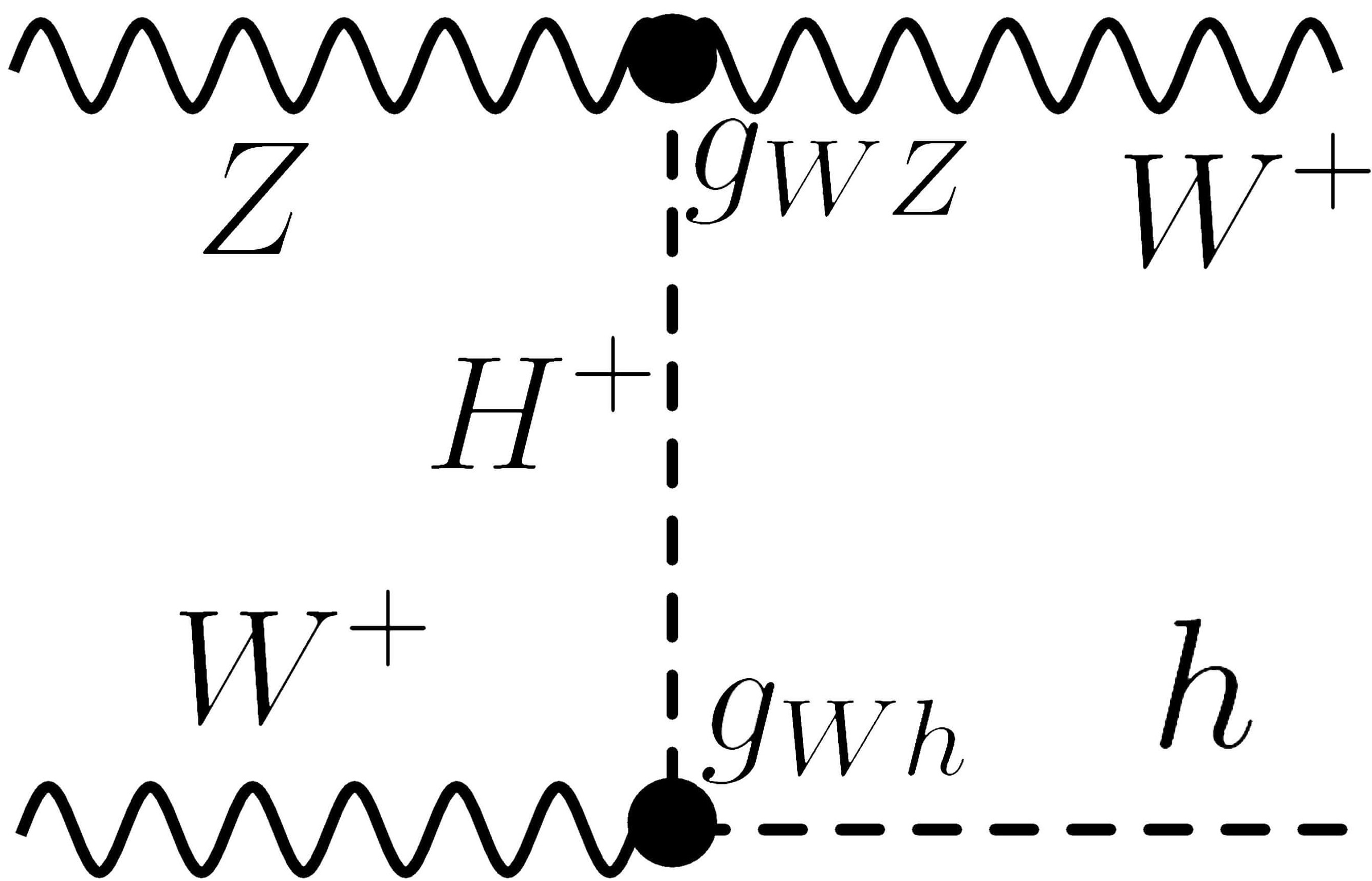}} \,  
 \resizebox{0.2\linewidth}{!}{ \includegraphics{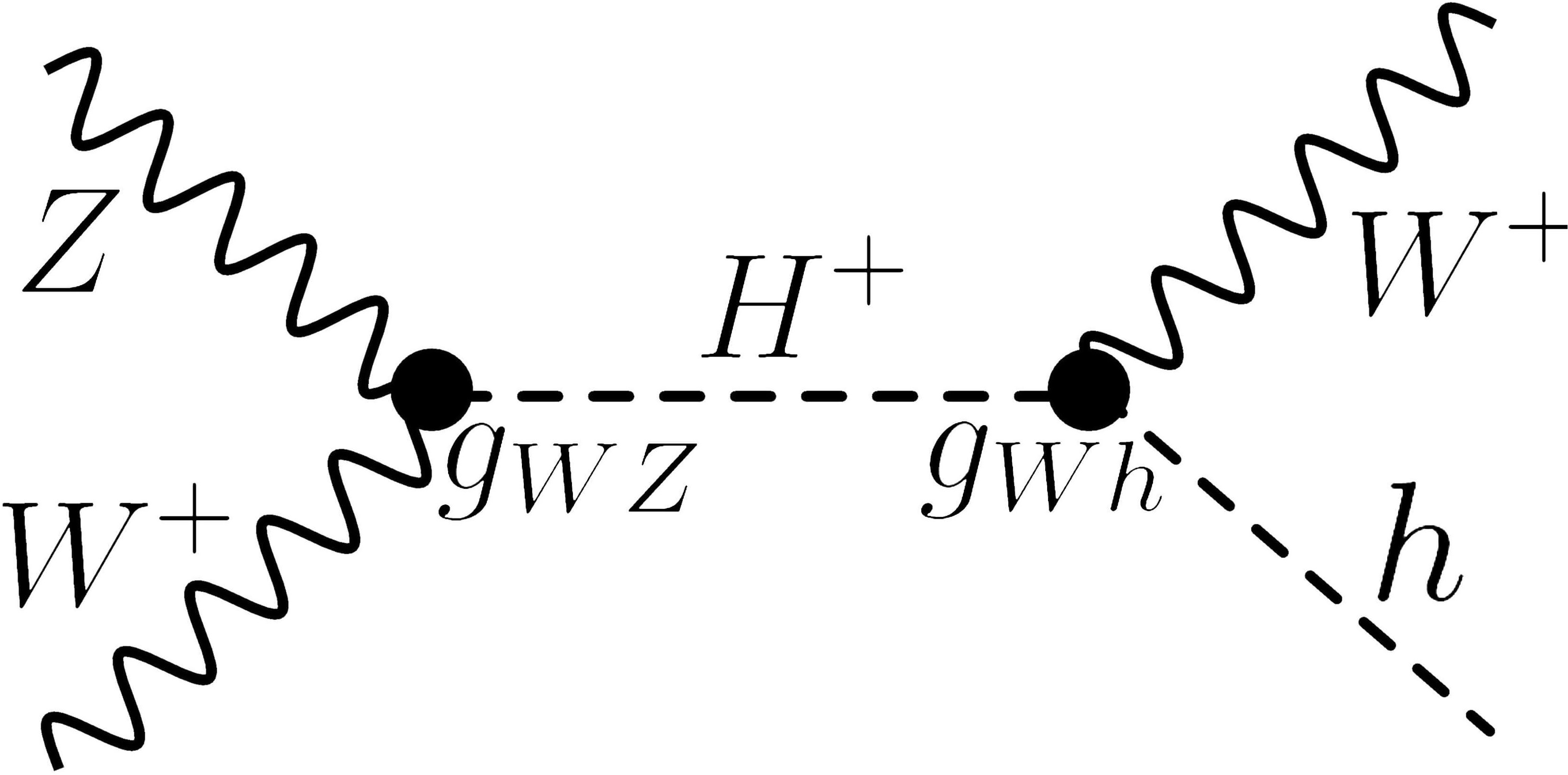}} 
 
 \vspace{0.5cm} 
 
 \resizebox{0.2\linewidth}{!}{ \includegraphics{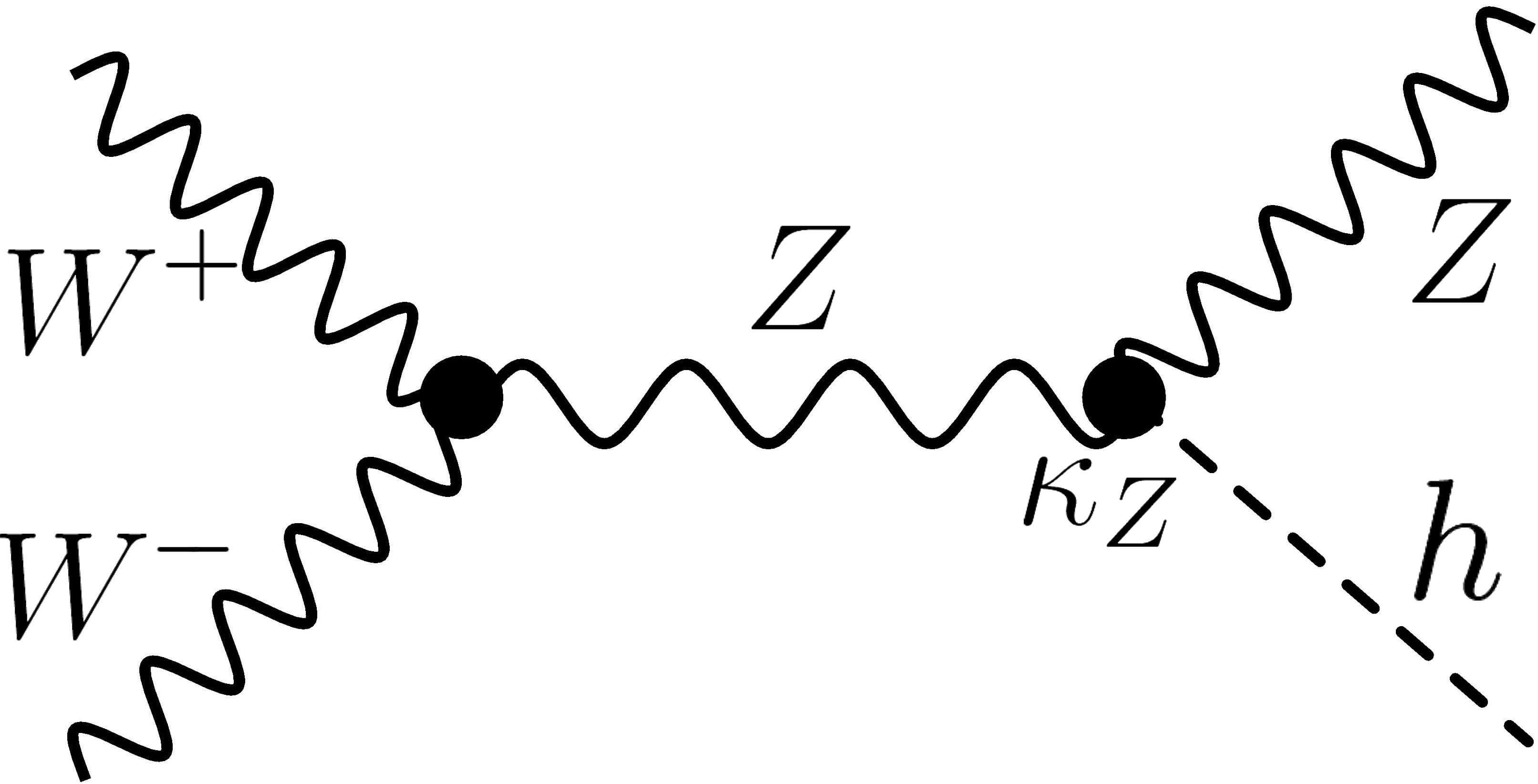}} \, \,  
 \resizebox{0.18\linewidth}{!}{ \includegraphics{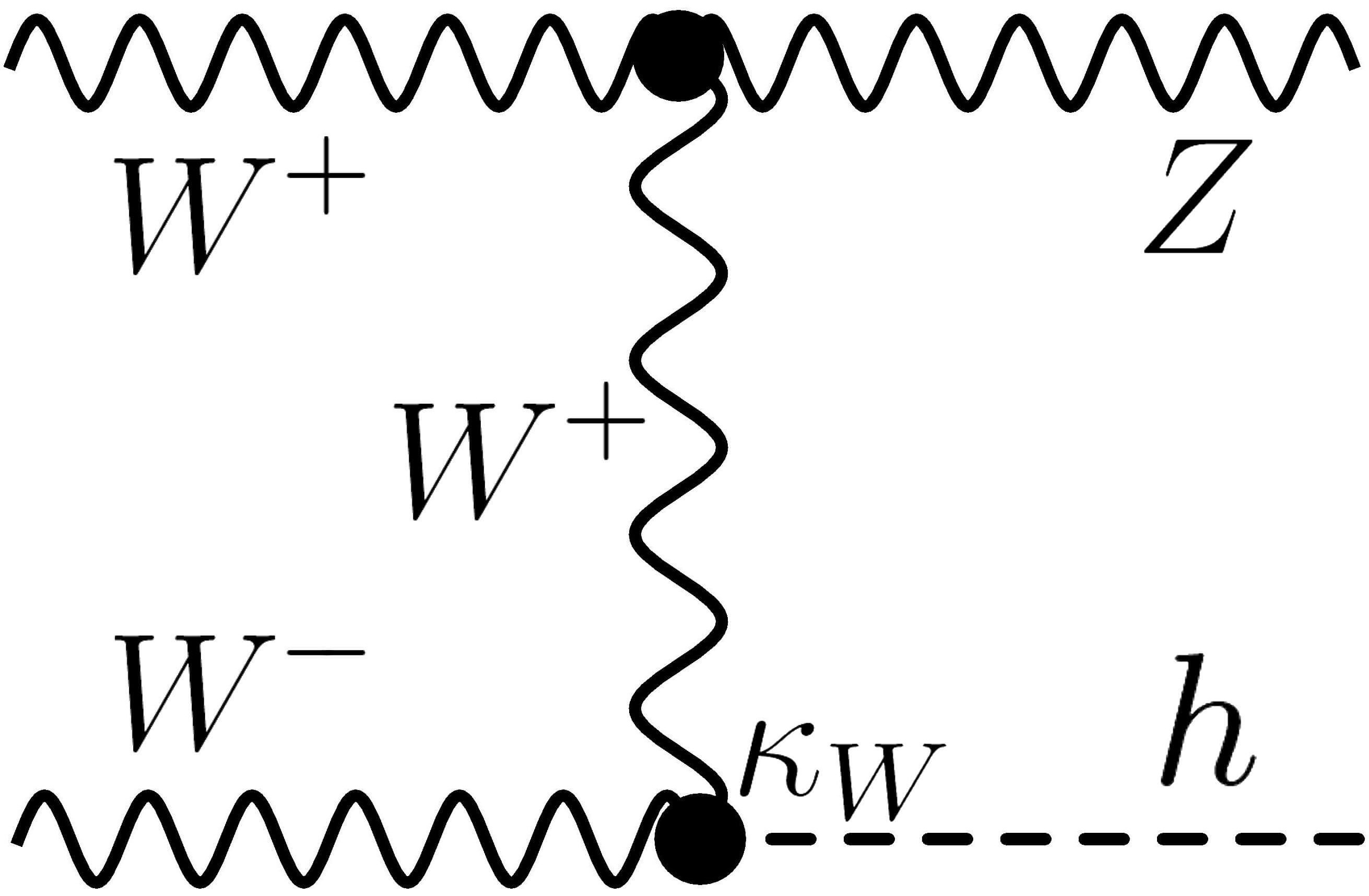}} \, \,  
 \resizebox{0.18\linewidth}{!}{ \includegraphics{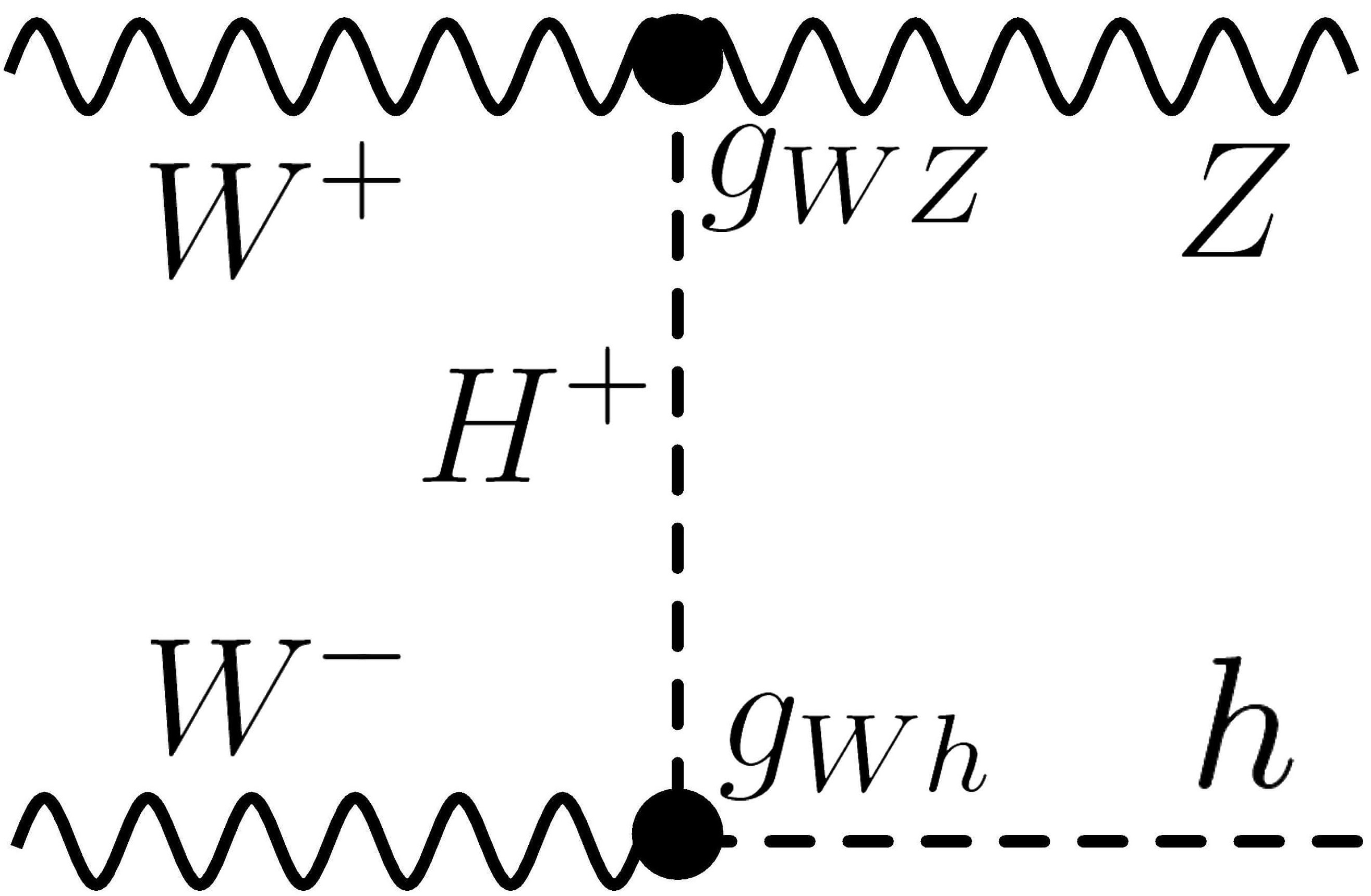}}  
\caption{Diagrams contributing to $W^\pm Z \rightarrow W^{\pm} h$ and $W^+ W^- \rightarrow Z h$, not including the diagrams related to those shown by exchange of final legs.}
\label{fig:ZhWHdiag}
\end{figure}

Let us work out how these new states enter for $W^{\pm}Z\rightarrow W^{\pm}h$ and $W^{+}W^{-}\rightarrow Zh$. The analysis is similar for both processes. The main difference between the two final states is the existence of an $s$-channel resonance for $W^{\pm}Z\rightarrow W^{\pm}h$. We can see the diagrams contributing to both processes in Figure~\ref{fig:ZhWHdiag}. As noted previously, the most substantial growth in energy occurs when all the gauge bosons are longitudinally polarized; we can write the amplitude for this polarization arrangement in the high energy limit as 
\begin{align}
    \mathcal{M}_{Zh}^{LLL} &= \frac{g^{2}\cos\theta}{4m_{W}^{2}} \left(\sum_{i}g^{i}_{WZ}g^{i}_{Wh} +\kappa_{W} -\kappa_{Z} \right) s + \mathcal{O}(s^0)\, , \\
     \mathcal{M}_{Wh}^{LLL} &= \frac{g^{2}(3+\cos\theta)}{8m_{W}^{2}} \left(\sum_{i}g^{i}_{WZ}g^{i}_{Wh} +\kappa_{W} -\kappa_{Z} \right) s  + \mathcal{O}(s^0)\,    ,
     \label{eq:mLLL}
\end{align}
where $\theta$ is the scattering angle in the center of mass frame, and $s$ is the usual Mandelstam variable. To prevent unitarity violation, any consistent weakly coupled extended Higgs sector must satisfy the following sum rule
\begin{align}\label{eq:unitirization}
\sum_{i}g^{i}_{WZ}g^{i}_{Wh} =\kappa_{Z} -\kappa_{W} \, . 
\end{align}
The new states tame the bad high-energy behavior but generate significant modifications at low energy. If, however, the mass of the lightest charged state is above the process energy, the description reduces to the one with only $\kappa$ modifiers as in~\cite{Stolarski:2020qim}. Notice that the sum rule is a universal feature of all perturbative UV completions.

\begin{figure}[t!]
  \resizebox{0.49\linewidth}{!}{ \includegraphics{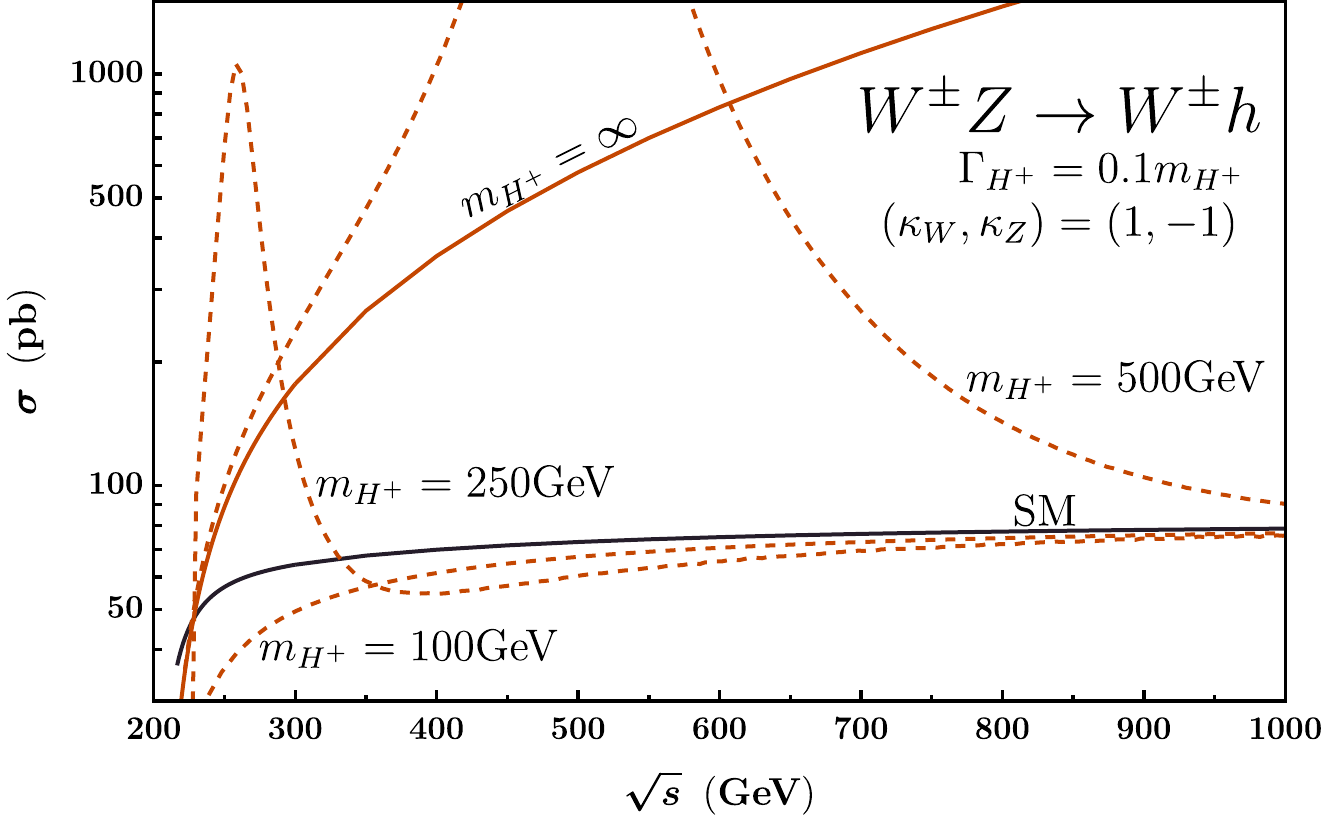}}\,`
 \resizebox{0.49\linewidth}{!}{ \includegraphics{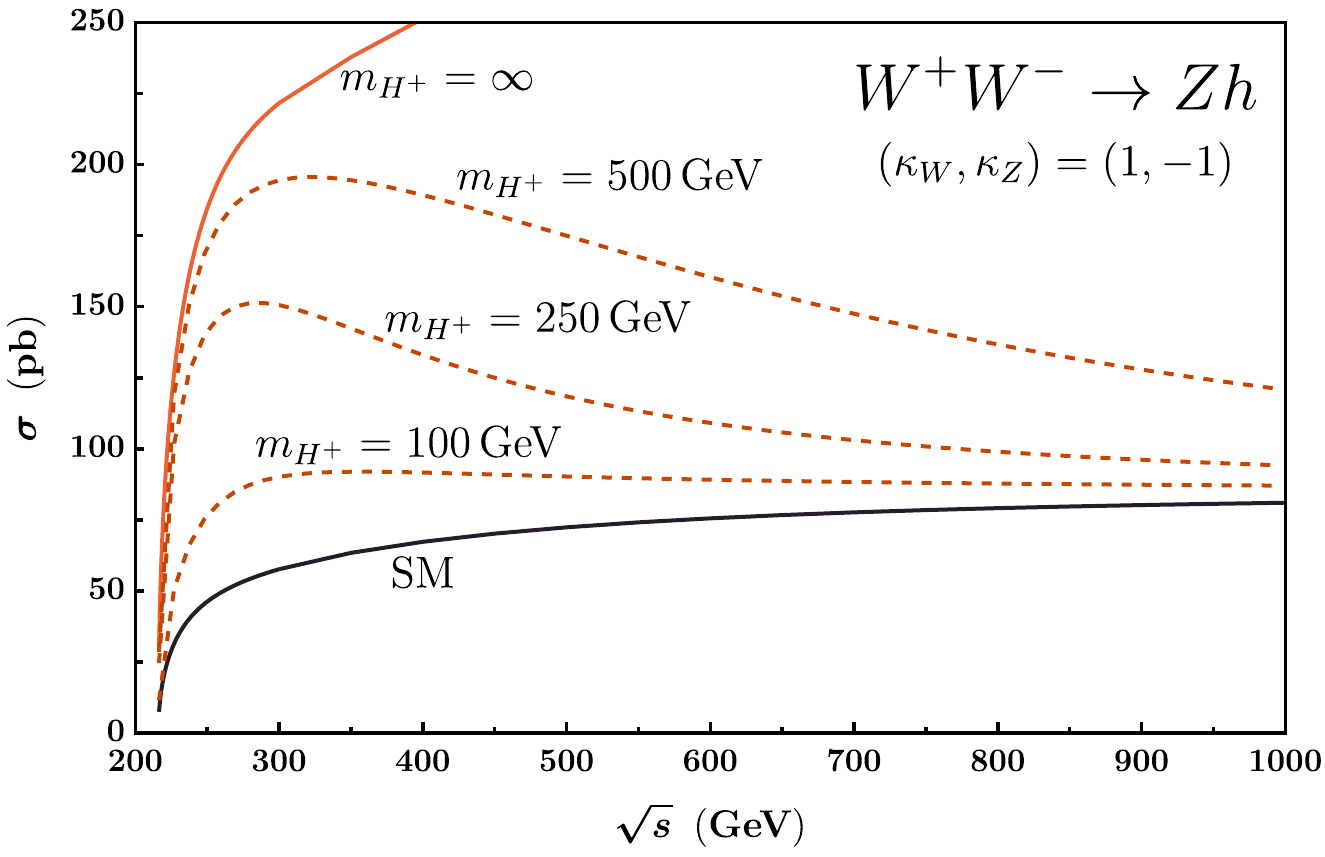}} 
\caption{Cross section of the $2 \rightarrow 2 $ process considering an effective description with one charged Higgs state with different masses, $\lambda_{WZ}=-1$ and the coupling fixed by unitarization [see Eq.~\eqref{eq:unitirization}]. Note that the left process is shown on a log scale, while the right is on a linear scale. }
\label{fig:crossP}
\end{figure}

We first analyze the case where there is only a single charged scalar. The cross section for different masses for both processes in the scenario of Eq.~\eqref{eq:wrong} with $\lambda_{WZ}=-1$ is shown in Figure~\ref{fig:crossP}. In the heavy mass limit, we recover the growth in energy of the longitudinal scattering, and we see that the cross section is huge. Notice the different scales in the processes, where the $Wh$ channel grows faster for intermediary energies than $Zh$. This was the feature exploited in~\cite{Stolarski:2020qim,ATLAS:2024vxc,CMS:2023sdc}. As we decrease the mass, the unitary restoration happens at lower energy, and at energies large compared to the mass, the cross section is comparable to the SM prediction. However, the differential cross section changes significantly, especially for scalar masses $\gtrsim 250$ GeV.

One crucial difference between the $W^{\pm}h$ and $Zh$ final state is the existence of an $s$-channel charged scalar resonance for the $W^{\pm} h$ final state.\footnote{If the additional neutral scalars conserve $CP$, there can be no $s$-channel neutral scalar exchange in the $WW\rightarrow Zh$ process. $CP$ is expected to be only an approximate symmetry~\cite{deLima:2024hnk} from the custodial-symmetry-breaking potential. Any induced $CP$-violating coefficients will be heavily suppressed and can be phenomenologically ignored.} For the plot and all the analysis, we consider a fixed width of $\Gamma_{H^{+}}=0.1 m_{H^{+}}$ as the low energy modifications scales with the inverse of width. The decay width is model-dependent, but every bound derived in this work gets stronger as we decrease the width, making this a conservative estimate. If the width were higher, this would indicate a breakdown in perturbativity for the specific model, and we would lose calculability. We also note in Figure~\ref{fig:crossP} that when the mass of the charged state is below the center of mass energy of the process, the $Z h$ channel shows a more substantial deviation from the SM.

\begin{figure}[t!]
  \resizebox{0.49\linewidth}{!}{ \includegraphics{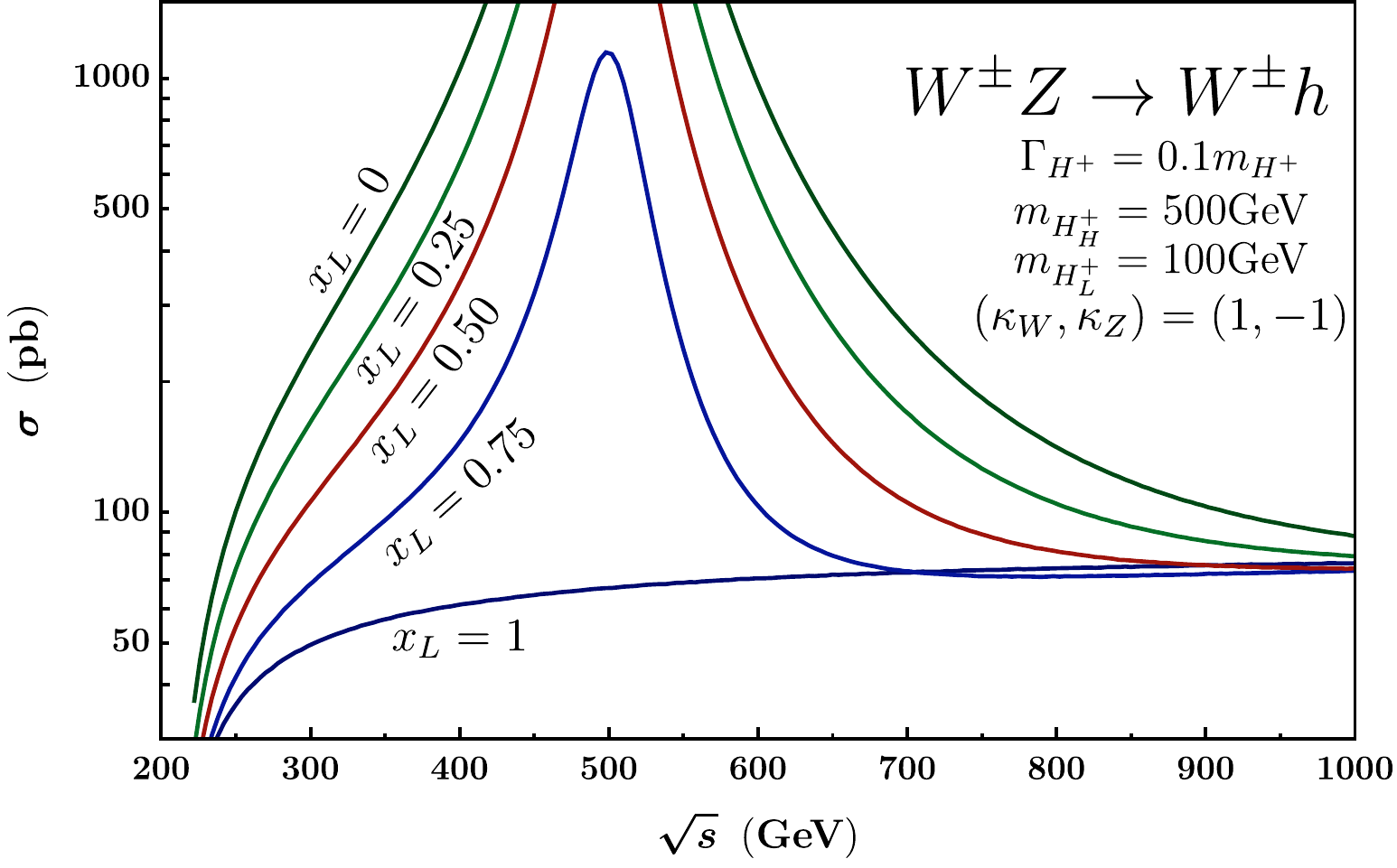}}\,
 \resizebox{0.49\linewidth}{!}{ \includegraphics{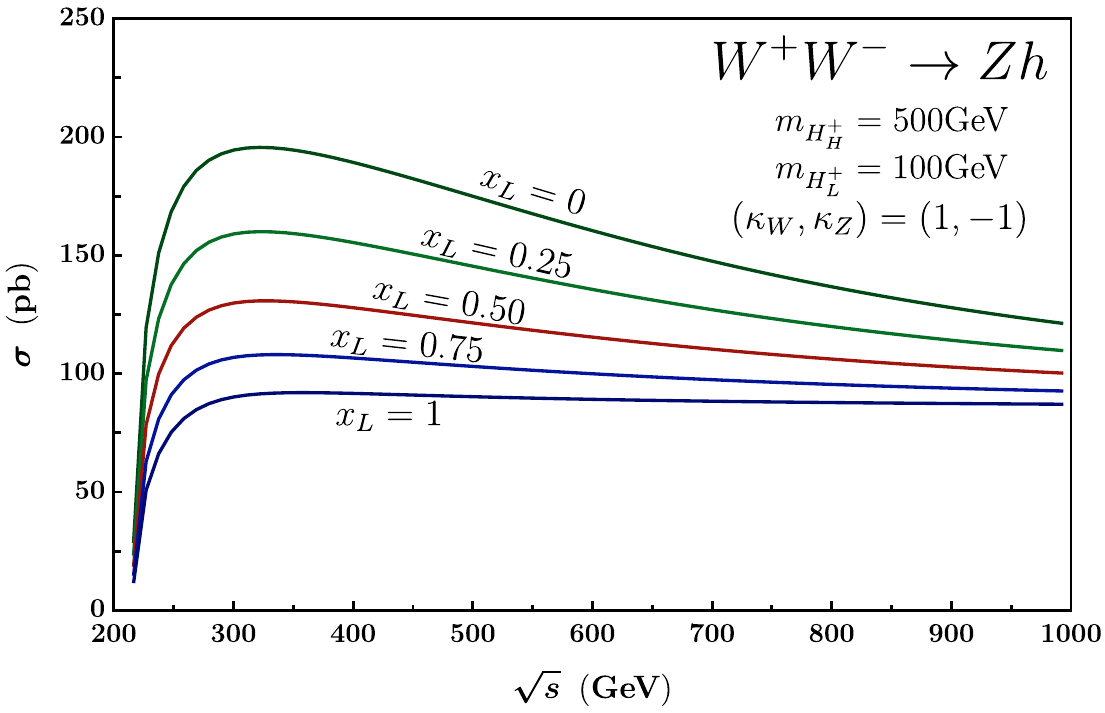}} 
\caption{Cross section of the $2 \rightarrow 2 $ processes considering two charged Higgs states, one with $m_{H} = 100~\text{GeV}$ and the other with $m_{H} = 500~\text{GeV}$ as a function of the fraction $x_{L}$ which quantifies the contribution of the lightest state to the unitarization with the coupling modifier fixed to $\lambda_{WZ}=-1$. The cases with $x_L=0$ and $x_L=1$ reduce to the one state analysis of Figure~\ref{fig:crossP}.}
\label{fig:crossTWO}
\end{figure}

In many scenarios, there will be multiple charged scalars, each contributing to the unitarity preservation. For example, if we have a set of $n$ scalars below the process energy and $k$ above, the matrix element will still grow with energy as in~Eq.~\eqref{eq:mLLL}, but the coefficient is smaller and proportional to 
\begin{align}
   \mathcal{M}^{LLL} \propto \left(\sum_{i = n+1}^{k} g^{i}_{WZ}g^{i}_{Wh} \right) s \, .
\end{align}
In this case, the growth in energy is reduced, and current bounds can be avoided. We introduce a fraction $x_{i}$ that each state contributes to the unitarization to deal with multiple states. This fraction introduces model-dependence, but one should expect to not rely on heavy fine-tuning. For example, if we have two charged states $H_{H}^{+}$ and $H_{L}^{+}$ the sum rule can be written as
\begin{align}
g^{L}_{WZ}g^{L}_{Wh} + g^{H}_{WZ}g^{H}_{Wh} = k_{Z}-k_{W} \, 
\end{align}
Since the sum is restricted, we can define the fractional contribution of a given state $x_i$ as
\begin{eqnarray}
x_{L}(\kappa_{Z}-\kappa_{W}) = g^{L}_{WZ}g^{L}_{Wh} \nonumber\\
 x_{H}(\kappa_{Z}-\kappa_{W}) = g^{H}_{WZ}g^{H}_{Wh}
 \label{eq:xdef}
\end{eqnarray}
with $x_{L}+x_{H}=1$. The cross section will be an interpolation between the theory with only a light state and a theory with only a heavy state, as shown in  Figure~\ref{fig:crossTWO}. The interpolation between the states is more evident in the $Zh$ channel. However, because of the resonance in the $Wh$ channel, the cross sections are all similar to the case with only one heavy state unless $|x_H| \ll 1$.

If the description with only one state is excluded for every mass value, then there is no way to hide the new physics by including more states. However, if a range of lighter masses is allowed, it is possible to weaken the bounds when the theory is tuned such that a lighter state is responsible for most of the unitarization. We discuss further this possibility in Appendix~\ref{ap}. In this case, there will still be a resonance, but the cross section will be substantially smaller than the one with only one state. It is generally difficult to justify such tuning from a UV picture where these coefficients come from diagonalizing a mass matrix. Note that the models we are considering are already significantly fine-tuned to set $\lambda_{WZ}=-1$.

When doing a collider analysis of this process, we expect the kinematics to have three distinct phases as a function of the charged scalar mass: when the scalar is lighter than the $Vh$ threshold, the kinematics are close to that of the SM. At large mass, the new states are approximately decoupled, and the process behaves similarly to the $\lambda_{WZ}=-1$ kinematics. There is an intermediate-mass regime with $200 \text{ GeV} \lesssim m_{H^+}\lesssim 400$ GeV where the kinematics are pretty different and poorly described by either limit. We can see these three regimes more explicitly in the recast of the ATLAS~\cite{ATLAS:2024vxc} analysis for $Wh$ done in the next section.

\section{ATLAS recast of VBF $Wh$}
\label{sec:ATLAS}
 \begin{figure}[t!]
  \resizebox{0.49\linewidth}{!}{ \includegraphics{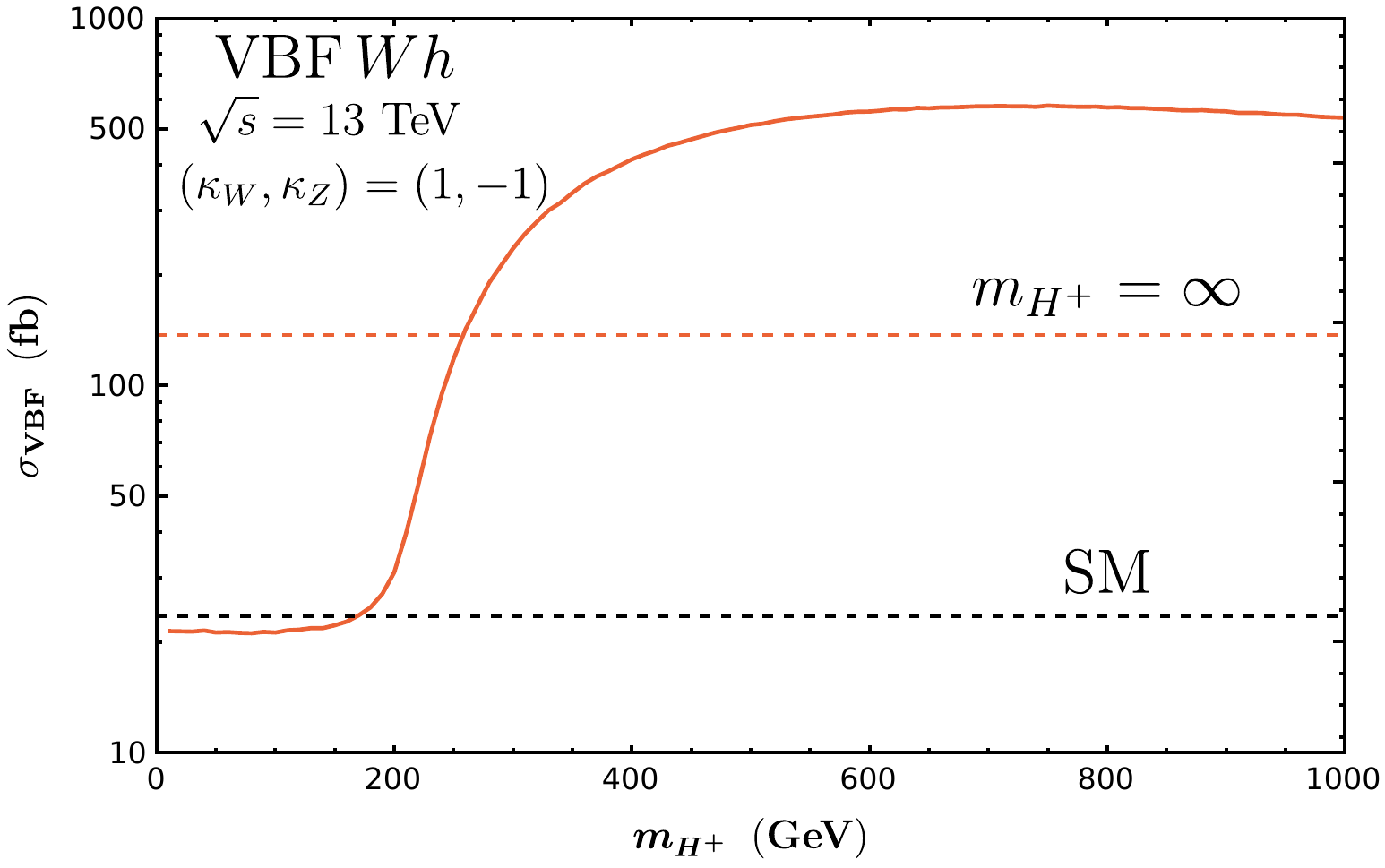}} 
 \resizebox{0.49\linewidth}{!}{ \includegraphics{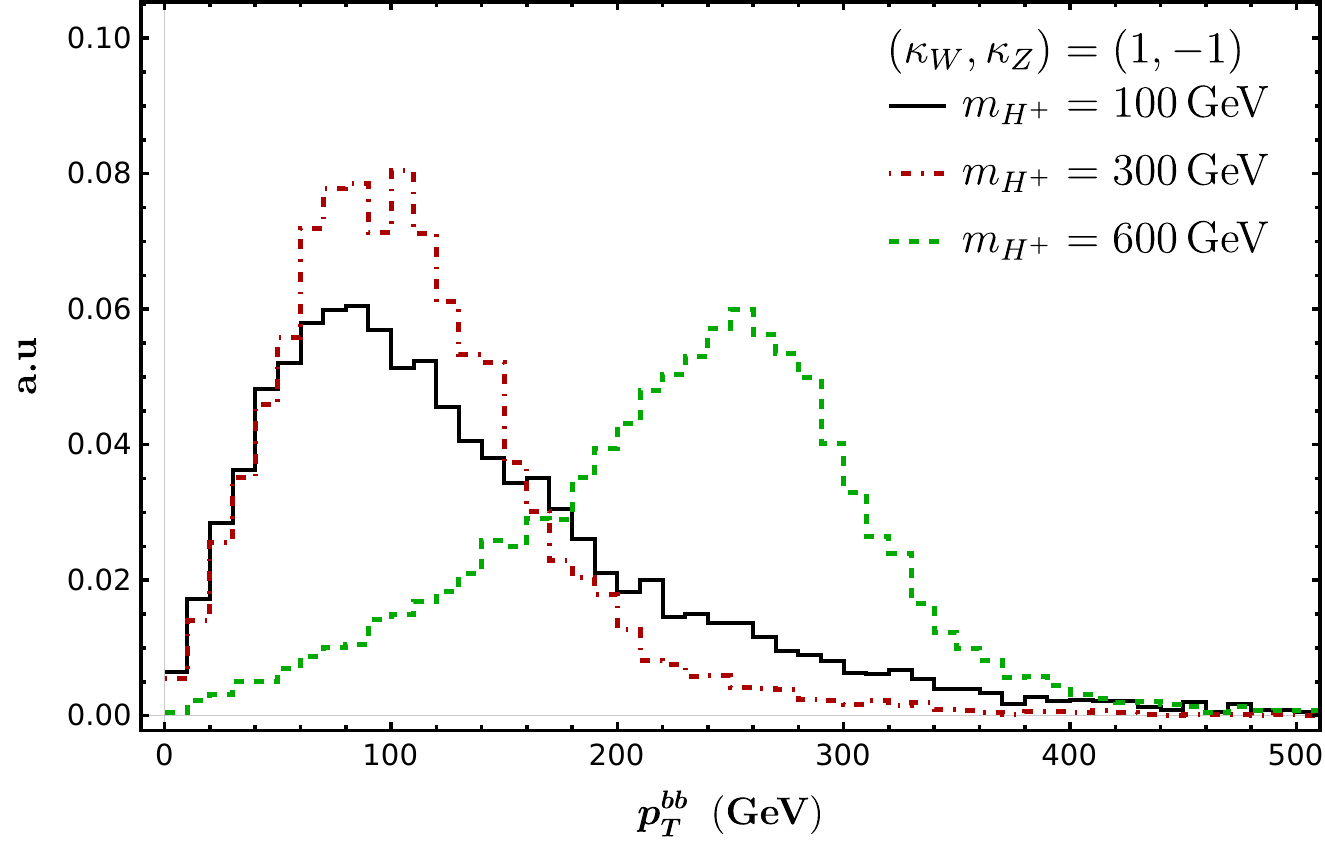}} 
\caption{
\textbf{Left:} Plot of the VBF cross-section at $\sqrt{s}= 13~\text{TeV}$ as a function of charged Higgs mass and fixed coupling modifier $\lambda_{WZ}=-1$. 
\textbf{Right:} Plot of the kinematic distribution at the same collision energy for the transverse momentum of the $b\bar{b}$ pair for different values of charged Higgs mass and fixed coupling modifier $\lambda_{WZ}=-1$. The SM distribution is similar to the one with $m_{H^{+}}=100 \text{ GeV}$. 
}
\label{fig:kin}
\end{figure}
In order to experimentally explore the $WZ \rightarrow Wh$ at the LHC, we use vector boson fusion (VBF), where the initial state vector bosons are radiated from the initial state quarks and fuse into the final state $Wh$. Following~\cite{Stolarski:2020qim,ATLAS:2024vxc,CMS:2023sdc}, we then require that the $W$ decays to $\ell \nu$ ($\ell = e,\mu$) and the $h$ decays to $b\bar{b}$.  
We simulate VBF $Wh$ using  MadGraph5\_aMC@NLO~\cite{Alwall:2014hca} using leading order accuracy in $\alpha_{s}$, interfaced to Pythia 8~\cite{Bierlich:2022pfr} for parton showering, hadronization and multiple parton interactions. We vary the mass of the charged state and fix its width to be $\Gamma_{H^+}=0.1m_{H^+}$. We perform the recast of the search performed in~\cite{ATLAS:2024vxc} using a RIVET~\cite{Bierlich:2019rhm} routine, giving us the acceptance for the three signal regions ($\text{SR}^{+}_{\text{loose}}$, $\text{SR}^{+}_{\text{tight}}$, $\text{SR}^{-}$) described in Table~\ref{tab:ATLAS}. We perform the analysis only for the scenario of Eq.~\eqref{eq:wrong} with $\lambda_{WZ}=-1$ as $\sim 10\%$ variations of that parameter consistent with the data will have a small effect. 

\begin{table*}[t!]
\centering
\caption{\label{tab:ATLAS} Kinematic cuts used for the three signal regions. The description of the kinematic variables can be found in Table 1 of~\cite{ATLAS:2024vxc}. }
\begin{tabular}{c|c|c|c}
\hline  & $\mathrm{SR}^{-}$ & $\mathrm{SR}_{\text {loose }}^{+}$ & $\mathrm{SR}_{\text {tight }}^{+}$ \\ \hline
$m_{b \bar{b}}$ & $\in(105,145) \mathrm{GeV}$ & $\in(105,145) \mathrm{GeV}$ & $\in(105,145) \mathrm{GeV}$ \\
$\Delta R_{b \bar{b}}$ &  $<1.2$ & $<1.6$ & $<1.2$ \\
$p_{\mathrm{T}}^{b \bar{b}}$ &   $>250 \mathrm{GeV}$ & $>100 \mathrm{GeV}$ & $>180 \mathrm{GeV}$ \\
$m_{j j}$ &  - & $>600 \mathrm{GeV}$ & $>1000 \mathrm{GeV}$ \\
$\Delta y_{j j}$ &    $>4.4$ & $>3.0$ & $>3.0$ \\
$m_{\text {top }}^{\text {lep }}$ &    $>260 \mathrm{GeV}$ & $>260 \mathrm{GeV}$  & $>260 \mathrm{GeV}$ \\
$\xi_{W b \bar{b}}$  &  $<0.3$ & $<0.3$ & $<0.3$ \\
$\Delta \phi(W b \bar{b}, j j)$ &   - & - & $>2.7$ \\ 
$N_{\text {jets }}^{\text {veto }}$ &  - & $\leq 1$ & $=0$ \\ \hline
\end{tabular}
\end{table*} 

{We show the cross section for VBF-$Wh$ at $\sqrt{s}= 13$ TeV in the left panel of Figure~\ref{fig:kin}. We first note that the VBF-VH cross-section with $\lambda_{WZ}=-1$ in the limit where $m_{H^+} \rightarrow \infty$ is much larger than the SM rate, which is the feature that was exploited in~\cite{Stolarski:2020qim}. On the other hand, when the charged Higgs is below threshold, $m_{H^+} \lesssim 200$ GeV, the cross section is very similar to the SM. The cross section increases significantly with increasing $m_{H^+}$, then slowly decreases above about 600 GeV. The cross section eventually asymptotes to the large $m_{H^+}$ limit, but the convergence is slow.

The acceptance in the signal regions depends on the kinematics, which, like the total cross section, depend strongly on $m_{H^+}$. We can see one example of kinematic dependence for different masses of the charged state in the right panel of Figure~\ref{fig:kin}, where we plot the normalized $p_T^{b\bar{b}}$ distribution. The search requires a minimum value of $p_T^{b\bar{b}}$, the transverse momentum of the $b\bar{b}$ system, because in the VBF topology, we expect the Higgs, which decays to $b\bar{b}$, is produced with substantial $p_T$. At low mass, the distribution looks similar to that of the SM (not shown in the figure). At intermediate masses, the location of the peak of the distribution is similar to the SM, but there are fewer events with high $p_T^{b\bar{b}}$. Finally, the distribution becomes concentrated at high $p_T^{b\bar{b}}$ at large masses, and the signal efficiency is expected to increase significantly. 

We estimate the acceptance using the above analysis chain and show our results on the left panel of Figure~\ref{fig:exc} as a function of the charged scalar mass. The figure shows the three distinct mass regions described above. When the charged scalar is below the $Wh$ threshold, the kinematics closely resemble the SM expectation; as the charged state can go on-shell, the kinematics change drastically, and the signal regions become much less efficient. Then, as we increase the mass further, the longitudinal scattering starts to dominate, and the kinematics become closely related to the analysis with only $\kappa$-modifiers in the negative coupling regime. It is possible to make the search more efficient in this intermediate region, but the gain is marginal, and we do not pursue such an improvement further in this work. 
 
 \begin{figure}[t!]
\resizebox{0.49\linewidth}{!}{ \includegraphics{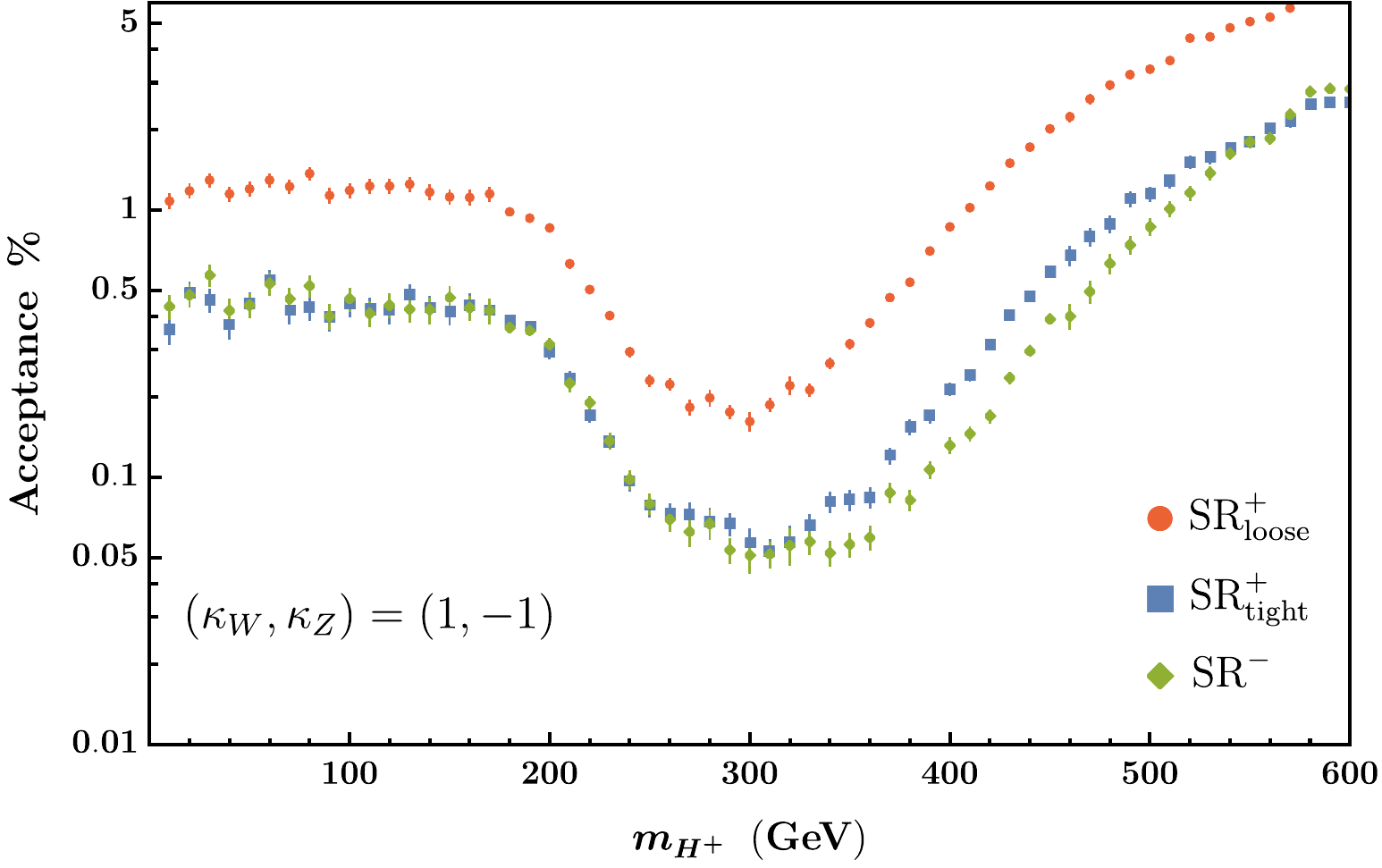}}\,
\resizebox{0.49\linewidth}{!}{ \includegraphics{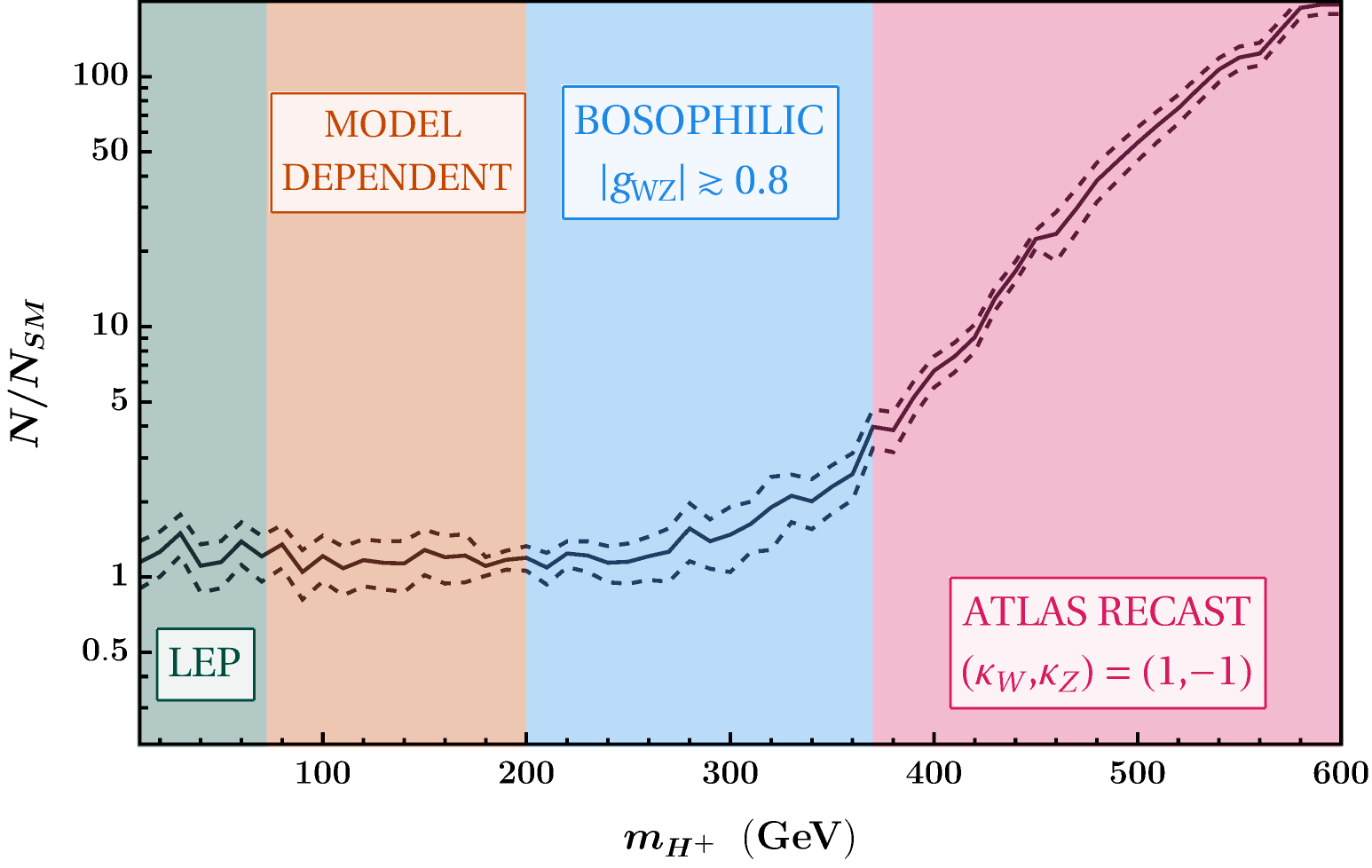}} 
\caption{
\textbf{Left:} Plot of the detector acceptance as a function of the charged scalar mass for the three signal regions $\text{SR}^{+}_{\text{loose}}$, $\text{SR}^{+}_{\text{tight}}$, $\text{SR}^{-}$ described in~\cite{ATLAS:2024vxc}. The Poisson uncertainty on the number of accepted events is the dominant one, as it is difficult to generate many events in MadGraph5\_aMC@NLO for certain parameter values. The asymptotic acceptance for small mass is the same as the SM value. 
\textbf{Right:} Ratio of the number of accepted events for a charged scalar model to the SM. This uses the signal region $\text{SR}^{-}$~\cite{ATLAS:2024vxc}. The dashed lines represent the $95\%$ CL Poisson uncertainty on the number of accepted events. This recast excludes charged scalar masses above $370 \text{ GeV}$ at $95\%$ CL. Below this mass, we include additional model-dependent searches that heavily restrict, but do not exclude, the negative coupling scenario. See Sec.~\ref{sec:other} for details. 
}
\label{fig:exc}
\end{figure}

 Assuming that the detector efficiency does not change much relative to the $\kappa$ analysis and using only ratios of events, we can use this analysis to put strong bounds on the scalar mass using the signal region optimized for the negative coupling modifier ($\text{SR}^{-}$). The high confidence exclusion from both  ATLAS~\cite{ATLAS:2024vxc} and CMS~\cite{CMS:2023sdc} can be understood as excluding the large mass regime. We recast the statistical analysis from ATLAS~\cite{ATLAS:2024vxc} by selecting a value of $\kappa_{W}$ and $\kappa_{Z}$ that lies on the boundary of their $95\%$ CL exclusion limit. Using this method, we can calculate the number of events corresponding to the boundary of the $95\%$ CL exclusion limit. We find that a signal sample larger than $N_{2\sigma}=3.94\times N_{SM}$ is excluded at $95\%$ CL. This can give a model-independent exclusion of masses larger than $370 ~\text{GeV}$ as shown in the right panel of Figure~\ref{fig:exc}. This exclusion of the mass should be interpreted as the mass of the lightest charged Higgs state, which is responsible for the unitarization. Including an allowed $10\%$ deviation of $\lambda_{WZ}$ does not significantly change the excluded mass window as the number of events grows quickly with increasing scalar mass. In Appendix~\ref{ap}, we show how it is possible to weaken this bound further by exploring the scenario with two charged Higgs states where the lightest is responsible for part of the unitarization.

 \section{Constraints From Other Processes}
\label{sec:other}
 
Since the current experimental searches cannot exclude the scenario with a charged scalar mass below 370 GeV, we explore other measurements that can constrain this scenario in this section. Any bound explored in this section will be model-dependent, only making UV completing the negative coupling hypothesis more difficult. In the next section, we point out the possibility of using the VBF $Zh$ channel to reach masses below 370 GeV in a model-independent way.

From the custodial violating interactions of the charged Higgs in Eq.\eqref{Eq:NPEQ}, we can see that, in the same way, that it contributes to $VV \rightarrow Vh $, it also contributes to $VV \rightarrow VV$. We can explore this scattering through the VBF $WZ$ channel, which will have scalar exchange diagrams similar to the first row of Figure~\ref{fig:ZhWHdiag}. For masses above the $WZ$ threshold, $m_{H^{+}} \gtrsim 170~\text{GeV}$, the resonant exchange can have a particularly large effect, and resonance searches in this channel have been performed by CMS~\cite{CMS:2021wlt}. This channel depends only on $g_{WZ}$ [see Eq.~\eqref{Eq:NPEQ}], while unitarization constrains the product $g_{WZ}\times g_{Wh}$ [see Eq.~\eqref{eq:unitirization}], so setting bounds becomes model-dependent. Probing the coupling $g_{Wh}$ independently is impossible experimentally as it is always tied to the charged Higgs decay, which can have other modes.
 
  If one assumes that in this mass region that decays to fermions are subdominant, then the dominant decay channels for the charged Higgs are $H^{\pm} \rightarrow W^{\pm}Z$ and $H^{\pm}\rightarrow W^{\pm}h$  (if kinematically allowed). In that case, it is possible to obtain exclusions for the charged Higgs mass, fixing the largest possible value of the $g_{WZ}$ coupling. In this approximation, we can see the effect of the branching ratio to $WZ$ as a function of $g_{WZ}$ in Figure~\ref{fig:gCMS}. We include this exclusion region in the right panel of Figure~\ref{fig:exc}, where we use the experimental $95\%$ CL bound on the on-shell production of the charged Higgs times the branching ratio to $W^{\pm}Z$ from~\cite{CMS:2021wlt} for $m_{H^{+}} > 200~\text{GeV}$. We can see the exact dependence on $g_{WZ}$ in the left panel Figure~\ref{fig:gCMS}. The search in~\cite{CMS:2021wlt} does not consider masses below 200 GeV. However, we see no reason why they would not have some sensitivity to lower mass states, and we encourage experimental groups to extend their searches to as low of mass as possible. 
  
 \begin{figure}[t!]
\resizebox{0.49\linewidth}{!}{ \includegraphics{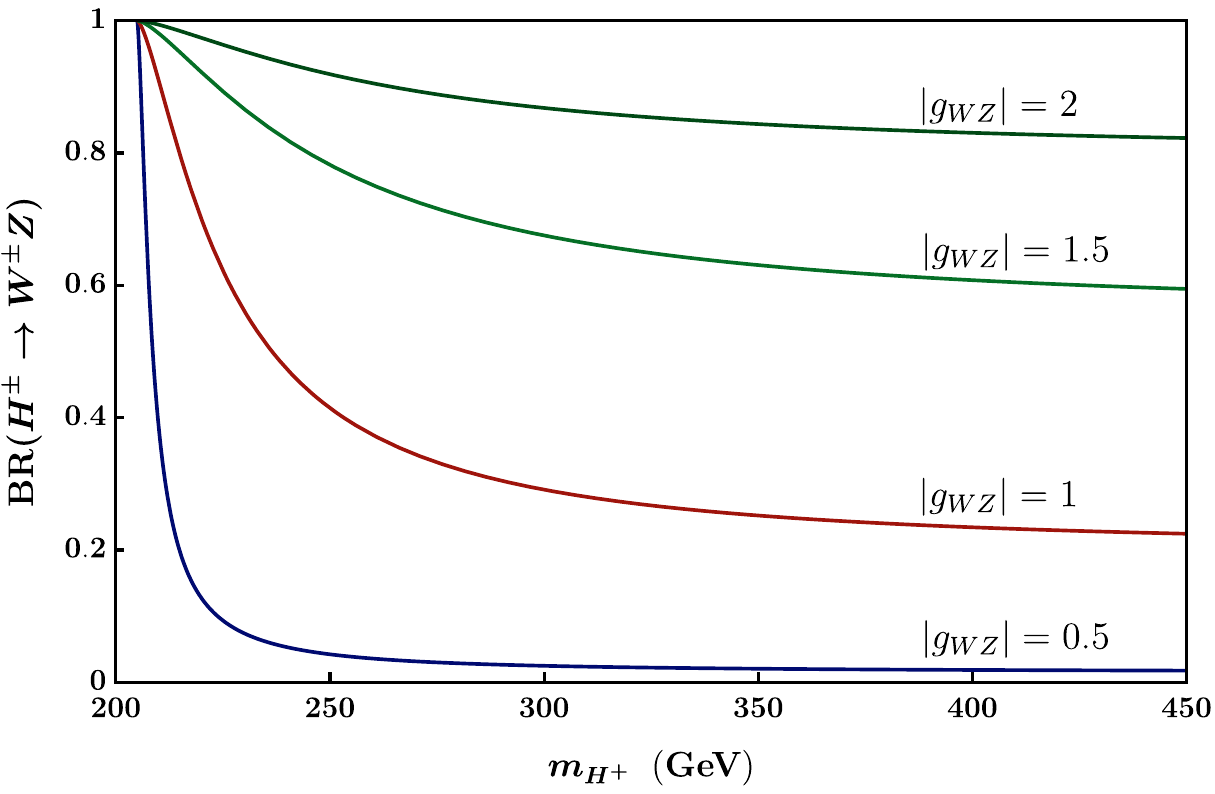}}\,
\resizebox{0.49\linewidth}{!}{ \includegraphics{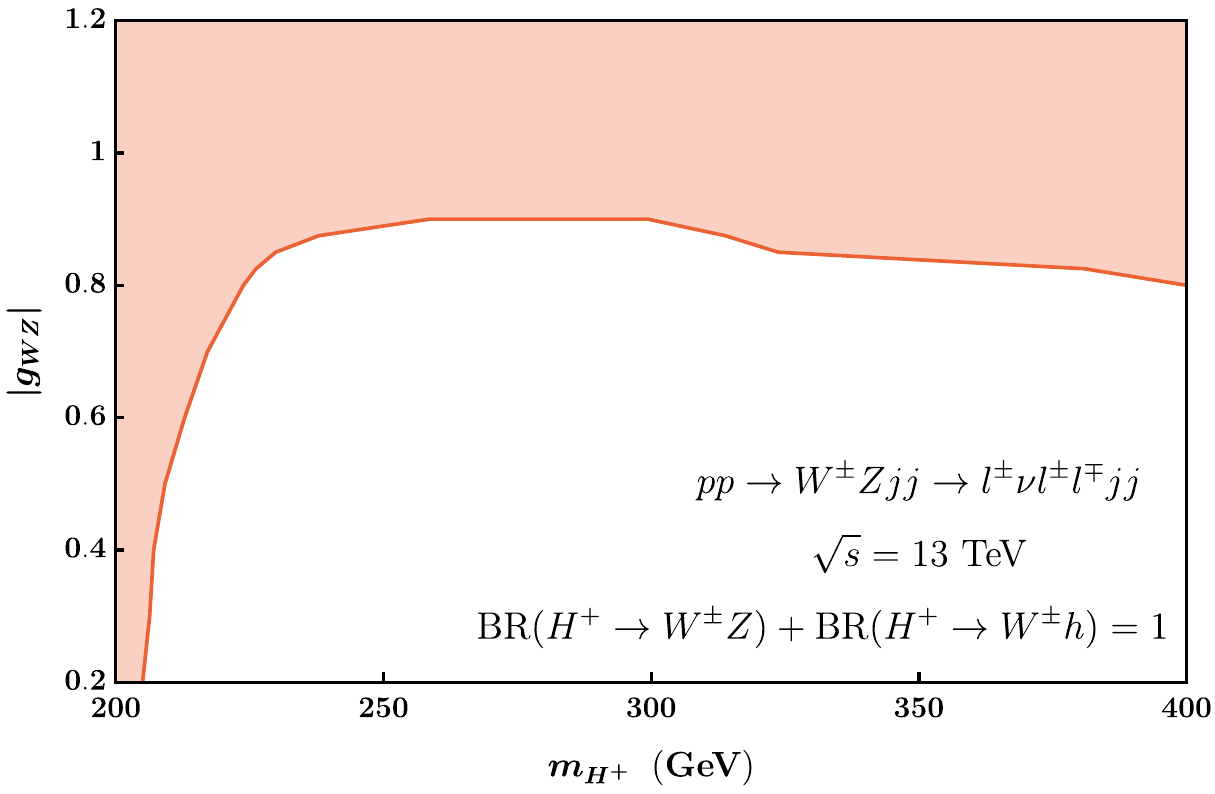}} 
\caption{
\textbf{Left:} Dependence on $g_{WZ}$ of the branching ratio to $WZ$, assuming no fermionic decay channels.
\textbf{Right:} Exclusion bound in the $g_{WZ}$ vs $m_{H^{+}}$ plane from~\cite{CMS:2021wlt}. Values above the curve are disfavored at $95\%$ CL.
}
\label{fig:gCMS}
\end{figure}  
  
We can use the unitarization condition of Eq.~\eqref{eq:unitirization} and the bound on $g_{WZ}$ above to place a bound on $g_{Wh}$. Numerically, the bounds on these two couplings for $m_{H^{+}} \gtrsim 225~\text{GeV}$ are:
\begin{align}
|g_{WZ}| < 0.8 \, , \, |g_{Wh}| > 2.5 \, ,
\end{align}
and are stronger going to lower masses down to $m_{H^{+}} = 200~\text{GeV}$. These bounds are weakened if there are substantial additional charged Higgs decay modes.

If the charged state does couple to fermions, there are multiple searches that place strong constraints. If the mass is below around $72.5~\text{GeV}$, we can obtain mostly model-independent\footnote{This bound assumes a fermionic coupling similar to the type-I two-Higgs-doublet model, which is the case for an arbitrary AC model.} generic bound on charged Higgs states from LEP~\cite{ALEPH:2013htx}. For larger masses, the searches depend heavily on the primary charged Higgs decay mode. In particular, if the dominant decay mode is $H^+\rightarrow \tau\nu$, it is possible to put strong bounds for masses above $m_{H^{+}} > 80$ GeV~\cite{CMS:2019bfg,ATLAS:2018gfm}. For a charged Higgs coupled to quarks, there are strong bounds in top decay searches $t\rightarrow H^+(cb) b$ for masses $m_{H^{+}}<172$ GeV~\cite{ATLAS:2023bzb}, or from  $H^+\rightarrow tb$ above $m_{H^{+}} > 200$ GeV~\cite{ATLAS:2021upq,CMS:2019rlz}.  

It is generally difficult to suppress the fermionic decays of the charged Higgs entirely, and therefore, if the charged Higgs is below the $WZ$ threshold, then decays to fermions are expected to dominate. If a fermiphobic charged Higgs could be engineered,  it may still be possible to put significant bounds in the parameter space from the off-shell weak decay $H^+ \rightarrow W^{+*}Z^*\rightarrow \ell\ell\ell\nu$ using the CMS multilepton search~\cite{CMS:2022nty}. 

As we mentioned, models that generate a negative gauge-Higgs couples inevitably have scalars in extended representations of $SU(2)_{L}$, which means the existence of neutral states and possibly states with electric charge larger than one. These states have distinct signatures~\cite{Ismail:2020kqz,Degrande:2017naf,PhysRevD.107.015018} that can be searched for. One important channel to look at comes from non-decoupling processes like $h \rightarrow \gamma \gamma$ and $h \rightarrow Z \gamma$. These channels usually give a strong bound on any specific UV completion, but these bounds are model-dependent. Thus, we cannot compute the bounds from these processes without specifying the entire model.

\section{Discussion}
\label{sec:conc}
\begin{figure}[t!]
 \resizebox{0.55\linewidth}{!}{ \includegraphics{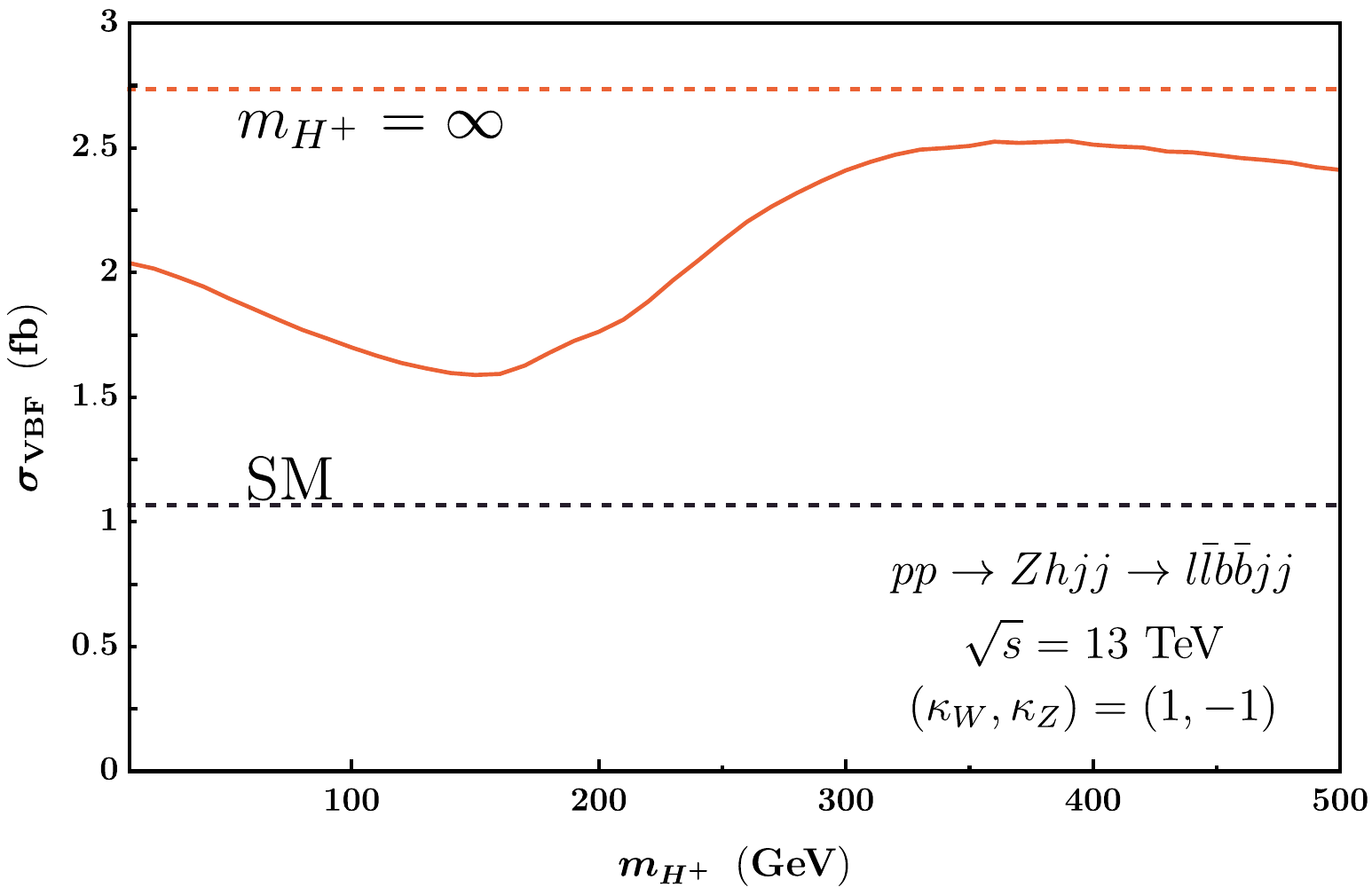}} 
\caption{Cross section times branching ratio for the VBF $Zh$ process with $\lambda_{WZ}=-1$ as a function of the mass of the charged Higgs. The cross section at high masses grows until reaching the $\lambda_{WZ}=-1$ value.}
\label{fig:VHVBF}
\end{figure}
This work explored the effects of charged scalar states beyond the Standard Model that contribute to the interference effects in $V \, V \rightarrow V \,  h $ processes. 
These processes are very sensitive to the sign of the gauge Higgs coupling~\cite{Stolarski:2020qim} via tree-level interference, and recent searches~\cite{ATLAS:2024vxc,CMS:2023sdc} claim an exclusion of the $\lambda_{WZ}=-1$ wrong-sign coupling regime with high confidence. We show that this high confidence exclusion becomes an exclusion of the high mass regime of any weakly coupled UV completion of the wrong sign scenario of Eq.~\eqref{eq:wrong}. We obtained a mostly model-independent bound on how heavy this new state can be and found that anything above $370 \text{ GeV}$ is experimentally excluded by a recast of the ATLAS analysis~\cite{ATLAS:2024vxc}. On the other hand, the scenario with $\lambda_{WZ}=-1$ and a charged scalar below 370 GeV cannot be excluded by our analysis. We expect the CMS results to be similar. We show in the appendix that we can further weaken this bound by including a light charged Higgs, which is mostly responsible for the unitarization of the longitudinal scattering.

If such a light-charged state exists, it could be probed by other processes. We discuss bounds from  $WZ$ scattering probed by VBF $WZ$, as well as bounds from fermionic decays of the charged Higgs that are complementary to the VBF $Wh$ search. The bounds from these channels are model-dependent and could be potentially evaded by clever model building and/or fine-tuning.  We note that several of these searches do not explore the entire range of masses they could be sensitive to, and we hope future iterations of these searches can expand the masses they explore. 
Two example models explored in~\cite{deLima:2021llm} are the AC pentet and AC sextet. These models that can avoid bounds from precision Higgs  data and the searches explored in this work. A complete analysis of explicit models is beyond the scope of this paper and is left for future work.

One potential direction to exclude the low mass regime in a model-independent way is to use the VBF $Zh$ process. It is vital to exclude all possible masses to ensure that the gauge-Higgs couplings are positive. This would mean that we are indeed in the decoupling regime in the Standard Model Higgs sector and that there is no large new physics contribution in this sector. For $\lambda_{WZ}=-1$, this channel has a significantly enhanced cross section relative to the SM even for very small charged Higgs masses, as shown in Figure~\ref{fig:VHVBF}. The enhancement factor is around 2 for masses below $370 \text{ GeV}$. Unfortunately, the signal in that channel is smaller than VBF $Wh$ by a factor of 20 when considering the relevant leptonic branching ratios. While the cross section is small, this process could perhaps be probed with HL-LHC or future lepton colliders. We leave this analysis for future work.

\acknowledgments
 We thank Dag Gillberg, John Keller, Heather Logan, Laura Miller, and Ezekiel Staats for their helpful conversations.
This work was supported in part by the Natural Sciences and Engineering Research Council of Canada (NSERC). C.H.L is supported by TRIUMF, which receives federal funding via a contribution agreement with the National Research Council (NRC) of Canada.

\appendix


\section{Two state ATLAS recast}
\label{ap}
\begin{figure}[b!]
 \resizebox{0.55\linewidth}{!}{ \includegraphics{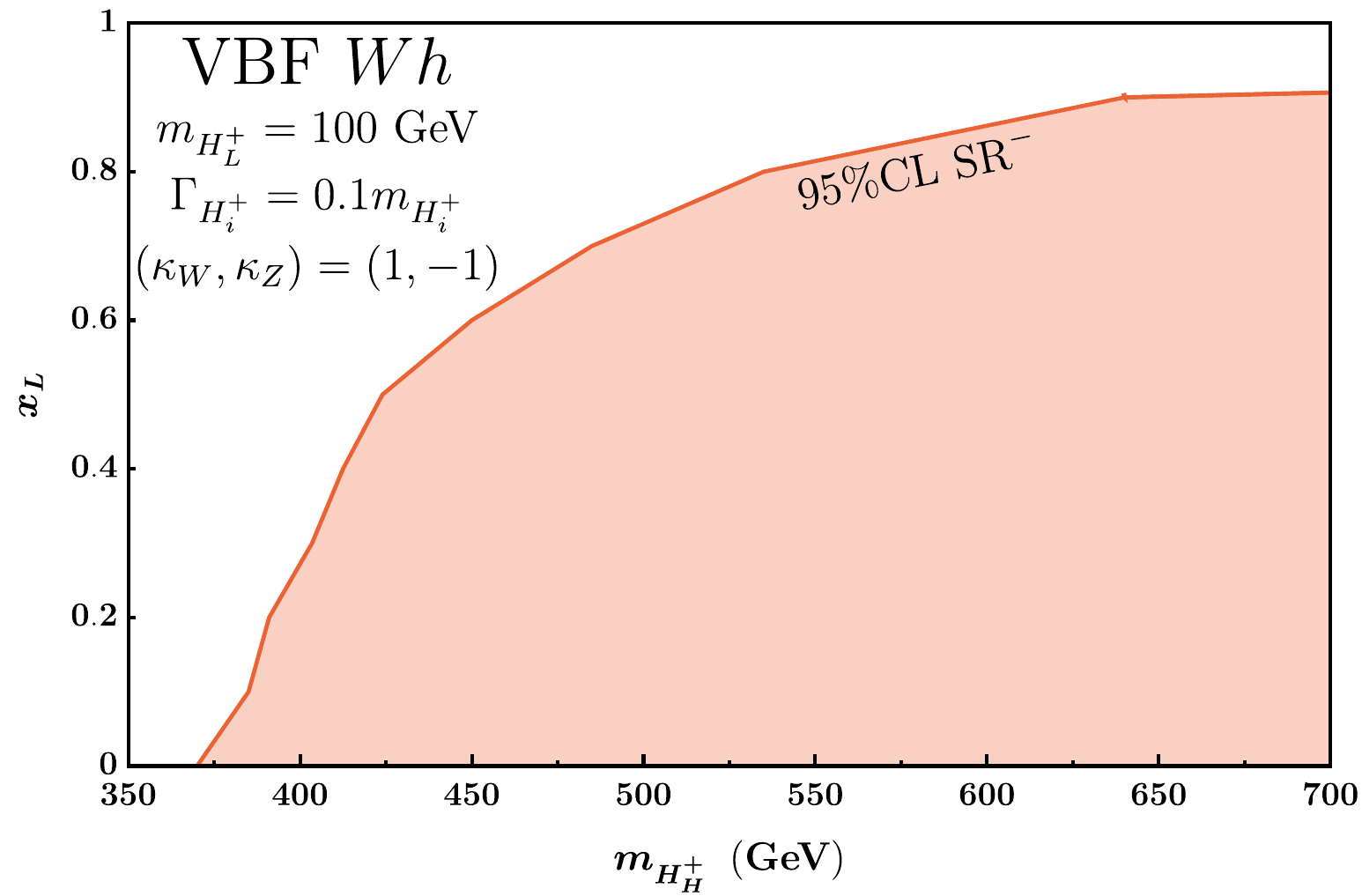}} 
\caption{Exclusion bounds based on the ATLAS VBF $Wh$ analysis for the two charged Higgs hypothesis and fixed coupling modifier $\lambda_{WZ}=-1$. We fix the mass of the lighter charged state to $100 \text{ GeV}$. We apply the same $95\%$ CL exclusion criterion as in Figure~\ref{fig:exc} where exclusion occurs if the number of events is larger than $3.94$ times the SM prediction using the $\text{SR}^{-}$ signal region. The excluded mass depends on the fraction of coupling of the light charged state, $x_{L}$. 
}
\label{fig:twoVBF}
\end{figure}
Here we further explore how the bounds can get weakened with multiple states if some of them are light. As a benchmark, we take a two state scenario and fix the mass of the lighter state to be 100 GeV. We can see in Figure~\ref{fig:crossP} that such a light state has little effect on the cross section of the $Wh$ final state. Following the parameterization defined in Eq.~\eqref{eq:xdef}, the strength of the coupling of the light charged Higgs is based on the fraction $x_{L}$ of the total unitarization coupling $(\kappa_{Z} -\kappa_{W})$ such that $x_{L}+x_{H}=1$. As $x_L$ gets larger and more of the coupling is in the light state, the heavy state's effect is hidden in the VBF $Wh$ process. 

We apply the $95\%$ CL exclusion criterion using the $\text{SR}^{-}$ signal region presented in Sec.~\ref{sec:ATLAS} and show the results in the plane of the heavy mass and $x_L$ in Figure~\ref{fig:twoVBF}. We see that for $x_L=0$, we reproduce the result of Figure~\ref{fig:exc}, but as $x_L$ increases, the bound gets weaker. This change is modest for $x_L \sim 0.5$, but if $x_L$ is tuned to be close to 1, the bound becomes much weaker. For example, for $x_{L}=0.95$ the bound from~\cite{ATLAS:2024vxc} only applies to masses above 1 TeV. The model-independent bound derived in the main paper is interpreted as the bound on the lightest charged state, which contributes to the unitarization, which for this benchmark of a light state with $m_{H^{+}_{L}} = 100~\text{GeV}$ is not excluded.

This analysis further motivates the exploration of VBF $Zh$ since low mass states can still have large effects on that process, as shown in Figure~\ref{fig:VHVBF}.

\bibliography{bibVhNP}

\begin{thebibliography}{40}
\expandafter\ifx\csname natexlab\endcsname\relax\def\natexlab#1{#1}\fi
\expandafter\ifx\csname bibnamefont\endcsname\relax
  \def\bibnamefont#1{#1}\fi
\expandafter\ifx\csname bibfnamefont\endcsname\relax
  \def\bibfnamefont#1{#1}\fi
\expandafter\ifx\csname citenamefont\endcsname\relax
  \def\citenamefont#1{#1}\fi
\expandafter\ifx\csname url\endcsname\relax
  \def\url#1{\texttt{#1}}\fi
\expandafter\ifx\csname urlprefix\endcsname\relax\def\urlprefix{URL }\fi
\providecommand{\bibinfo}[2]{#2}
\providecommand{\eprint}[2][]{\url{#2}}

\bibitem[{\citenamefont{Lee et~al.}(1977{\natexlab{a}})\citenamefont{Lee,
  Quigg, and Thacker}}]{Lee:1977yc}
\bibinfo{author}{\bibfnamefont{B.~W.} \bibnamefont{Lee}},
  \bibinfo{author}{\bibfnamefont{C.}~\bibnamefont{Quigg}}, \bibnamefont{and}
  \bibinfo{author}{\bibfnamefont{H.~B.} \bibnamefont{Thacker}}, ``{The Strength
  of Weak Interactions at Very High-Energies and the Higgs Boson Mass},''
  \bibinfo{journal}{Phys. Rev. Lett.} \textbf{\bibinfo{volume}{38}},
  \bibinfo{pages}{883} (\bibinfo{year}{1977}{\natexlab{a}}).

\bibitem[{\citenamefont{Lee et~al.}(1977{\natexlab{b}})\citenamefont{Lee,
  Quigg, and Thacker}}]{Lee:1977eg}
\bibinfo{author}{\bibfnamefont{B.~W.} \bibnamefont{Lee}},
  \bibinfo{author}{\bibfnamefont{C.}~\bibnamefont{Quigg}}, \bibnamefont{and}
  \bibinfo{author}{\bibfnamefont{H.~B.} \bibnamefont{Thacker}}, ``{Weak
  Interactions at Very High-Energies: The Role of the Higgs Boson Mass},''
  \bibinfo{journal}{Phys. Rev. D} \textbf{\bibinfo{volume}{16}},
  \bibinfo{pages}{1519} (\bibinfo{year}{1977}{\natexlab{b}}).

\bibitem[{\citenamefont{Logan}(2022)}]{Logan:2022uus}
\bibinfo{author}{\bibfnamefont{H.~E.} \bibnamefont{Logan}}, ``{Lectures on
  perturbative unitarity and decoupling in Higgs physics},''
  (\bibinfo{year}{2022}), \eprint{2207.01064}.

\bibitem[{\citenamefont{Tumasyan et~al.}(2022{\natexlab{a}})}]{CMS:2022dwd}
\bibinfo{author}{\bibfnamefont{A.}~\bibnamefont{Tumasyan}} \bibnamefont{et~al.}
  (\bibinfo{collaboration}{CMS}), ``{A portrait of the Higgs boson by the CMS
  experiment ten years after the discovery.},'' \bibinfo{journal}{Nature}
  \textbf{\bibinfo{volume}{607}}, \bibinfo{pages}{60}
  (\bibinfo{year}{2022}{\natexlab{a}}), \eprint{2207.00043}.

\bibitem[{\citenamefont{Aad et~al.}(2022)}]{ATLAS:2022vkf}
\bibinfo{author}{\bibfnamefont{G.}~\bibnamefont{Aad}} \bibnamefont{et~al.}
  (\bibinfo{collaboration}{ATLAS}), ``{A detailed map of Higgs boson
  interactions by the ATLAS experiment ten years after the discovery},''
  \bibinfo{journal}{Nature} \textbf{\bibinfo{volume}{607}}, \bibinfo{pages}{52}
  (\bibinfo{year}{2022}), \bibinfo{note}{[Erratum: Nature 612, E24 (2022)]},
  \eprint{2207.00092}.

\bibitem[{\citenamefont{Chen et~al.}(2016)\citenamefont{Chen, Lykken,
  Spiropulu, Stolarski, and Vega-Morales}}]{Chen:2016ofc}
\bibinfo{author}{\bibfnamefont{Y.}~\bibnamefont{Chen}},
  \bibinfo{author}{\bibfnamefont{J.}~\bibnamefont{Lykken}},
  \bibinfo{author}{\bibfnamefont{M.}~\bibnamefont{Spiropulu}},
  \bibinfo{author}{\bibfnamefont{D.}~\bibnamefont{Stolarski}},
  \bibnamefont{and}
  \bibinfo{author}{\bibfnamefont{R.}~\bibnamefont{Vega-Morales}}, ``{Golden
  Probe of Electroweak Symmetry Breaking},'' \bibinfo{journal}{Phys. Rev.
  Lett.} \textbf{\bibinfo{volume}{117}}, \bibinfo{pages}{241801}
  (\bibinfo{year}{2016}), \eprint{1608.02159}.

\bibitem[{\citenamefont{Chiang et~al.}(2018)\citenamefont{Chiang, He, and
  Li}}]{Chiang:2018fqf}
\bibinfo{author}{\bibfnamefont{C.-W.} \bibnamefont{Chiang}},
  \bibinfo{author}{\bibfnamefont{X.-G.} \bibnamefont{He}}, \bibnamefont{and}
  \bibinfo{author}{\bibfnamefont{G.}~\bibnamefont{Li}}, ``{Measuring the ratio
  of HWW and HZZ couplings through WWH production},'' \bibinfo{journal}{JHEP}
  \textbf{\bibinfo{volume}{08}}, \bibinfo{pages}{126} (\bibinfo{year}{2018}),
  \eprint{1805.01689}.

\bibitem[{\citenamefont{Stolarski and Wu}(2020)}]{Stolarski:2020qim}
\bibinfo{author}{\bibfnamefont{D.}~\bibnamefont{Stolarski}} \bibnamefont{and}
  \bibinfo{author}{\bibfnamefont{Y.}~\bibnamefont{Wu}}, ``{Tree-level
  interference in vector boson fusion production of Vh},''
  \bibinfo{journal}{Phys. Rev. D} \textbf{\bibinfo{volume}{102}},
  \bibinfo{pages}{033006} (\bibinfo{year}{2020}), \eprint{2006.09374}.

\bibitem[{\citenamefont{Xie and Yan}(2021)}]{Xie:2021xtl}
\bibinfo{author}{\bibfnamefont{K.-P.} \bibnamefont{Xie}} \bibnamefont{and}
  \bibinfo{author}{\bibfnamefont{B.}~\bibnamefont{Yan}}, ``{Probing the
  electroweak symmetry breaking with Higgs production at the LHC},''
  \bibinfo{journal}{Phys. Lett. B} \textbf{\bibinfo{volume}{820}},
  \bibinfo{pages}{136515} (\bibinfo{year}{2021}), \eprint{2104.12689}.

\bibitem[{\citenamefont{Das et~al.}(2024)\citenamefont{Das, Kundu, Levy,
  Prasad, Saha, and Sarkar}}]{Das:2024xre}
\bibinfo{author}{\bibfnamefont{D.}~\bibnamefont{Das}},
  \bibinfo{author}{\bibfnamefont{A.}~\bibnamefont{Kundu}},
  \bibinfo{author}{\bibfnamefont{M.}~\bibnamefont{Levy}},
  \bibinfo{author}{\bibfnamefont{A.~M.} \bibnamefont{Prasad}},
  \bibinfo{author}{\bibfnamefont{I.}~\bibnamefont{Saha}}, \bibnamefont{and}
  \bibinfo{author}{\bibfnamefont{A.}~\bibnamefont{Sarkar}}, ``{Sign of the
  $hZZ$ coupling and implication for new physics},''  (\bibinfo{year}{2024}),
  \eprint{2402.09352}.

\bibitem[{\citenamefont{Aad et~al.}(2024)}]{ATLAS:2024vxc}
\bibinfo{author}{\bibfnamefont{G.}~\bibnamefont{Aad}} \bibnamefont{et~al.}
  (\bibinfo{collaboration}{ATLAS}), ``{Determination of the relative sign of
  the Higgs boson couplings to $W$ and $Z$ bosons using $WH$ production via
  vector-boson fusion with the ATLAS detector},''  (\bibinfo{year}{2024}),
  \eprint{2402.00426}.

\bibitem[{\citenamefont{\text{CMS Collaboration}}(2023)}]{CMS:2023sdc}
\bibinfo{author}{\bibnamefont{\text{CMS Collaboration}}}
  (\bibinfo{collaboration}{CMS}), ``{Search for anomalous Higgs boson couplings
  in WH$\to\ell\nu\mathrm{b\overline{b}}$ production through Vector Boson
  Scattering},''  (\bibinfo{year}{2023}),
  \urlprefix\url{https://cds.cern.ch/record/2882655}.

\bibitem[{\citenamefont{Veltman}(1977)}]{Veltman:1977kh}
\bibinfo{author}{\bibfnamefont{M.~J.~G.} \bibnamefont{Veltman}}, ``{Limit on
  Mass Differences in the Weinberg Model},'' \bibinfo{journal}{Nucl. Phys. B}
  \textbf{\bibinfo{volume}{123}}, \bibinfo{pages}{89} (\bibinfo{year}{1977}).

\bibitem[{\citenamefont{Gunion et~al.}(2000)\citenamefont{Gunion, Haber, Kane,
  and Dawson}}]{Gunion:1989we}
\bibinfo{author}{\bibfnamefont{J.~F.} \bibnamefont{Gunion}},
  \bibinfo{author}{\bibfnamefont{H.~E.} \bibnamefont{Haber}},
  \bibinfo{author}{\bibfnamefont{G.~L.} \bibnamefont{Kane}}, \bibnamefont{and}
  \bibinfo{author}{\bibfnamefont{S.}~\bibnamefont{Dawson}},
  \emph{\bibinfo{title}{{The Higgs Hunter's Guide}}}, vol.~\bibinfo{volume}{80}
  (\bibinfo{year}{2000}).

\bibitem[{\citenamefont{de~Lima and Logan}(2022)}]{deLima:2022yvn}
\bibinfo{author}{\bibfnamefont{C.~H.} \bibnamefont{de~Lima}} \bibnamefont{and}
  \bibinfo{author}{\bibfnamefont{H.~E.} \bibnamefont{Logan}}, ``{Unavoidable
  Higgs coupling deviations in the Z2-symmetric Georgi-Machacek model},''
  \bibinfo{journal}{Phys. Rev. D} \textbf{\bibinfo{volume}{106}},
  \bibinfo{pages}{115020} (\bibinfo{year}{2022}), \eprint{2209.08393}.

\bibitem[{\citenamefont{Zyla et~al.}(2020)}]{ParticleDataGroup:2020ssz}
\bibinfo{author}{\bibfnamefont{P.~A.} \bibnamefont{Zyla}} \bibnamefont{et~al.}
  (\bibinfo{collaboration}{Particle Data Group}), ``{Review of Particle
  Physics},'' \bibinfo{journal}{PTEP} \textbf{\bibinfo{volume}{2020}},
  \bibinfo{pages}{083C01} (\bibinfo{year}{2020}).

\bibitem[{\citenamefont{de~Lima et~al.}(2022)\citenamefont{de~Lima, Stolarski,
  and Wu}}]{deLima:2021llm}
\bibinfo{author}{\bibfnamefont{C.~H.} \bibnamefont{de~Lima}},
  \bibinfo{author}{\bibfnamefont{D.}~\bibnamefont{Stolarski}},
  \bibnamefont{and} \bibinfo{author}{\bibfnamefont{Y.}~\bibnamefont{Wu}},
  ``{Status of negative coupling modifiers for extended Higgs sectors},''
  \bibinfo{journal}{Phys. Rev. D} \textbf{\bibinfo{volume}{105}},
  \bibinfo{pages}{035019} (\bibinfo{year}{2022}), \bibinfo{note}{[Erratum:
  Phys.Rev.D 108, 099901 (2023)]}, \eprint{2111.02533}.

\bibitem[{\citenamefont{Logan and Rentala}(2015)}]{GGM}
\bibinfo{author}{\bibfnamefont{H.~E.} \bibnamefont{Logan}} \bibnamefont{and}
  \bibinfo{author}{\bibfnamefont{V.}~\bibnamefont{Rentala}}, ``{All the
  generalized Georgi-Machacek models},'' \bibinfo{journal}{Phys. Rev. D}
  \textbf{\bibinfo{volume}{92}}, \bibinfo{pages}{075011}
  (\bibinfo{year}{2015}), \eprint{1502.01275}.

\bibitem[{\citenamefont{Gunion et~al.}(1991)\citenamefont{Gunion, Haber, and
  Wudka}}]{Gunion:1990kf}
\bibinfo{author}{\bibfnamefont{J.~F.} \bibnamefont{Gunion}},
  \bibinfo{author}{\bibfnamefont{H.~E.} \bibnamefont{Haber}}, \bibnamefont{and}
  \bibinfo{author}{\bibfnamefont{J.}~\bibnamefont{Wudka}}, ``{Sum rules for
  Higgs bosons},'' \bibinfo{journal}{Phys. Rev. D}
  \textbf{\bibinfo{volume}{43}}, \bibinfo{pages}{904} (\bibinfo{year}{1991}).

\bibitem[{\citenamefont{Belvedere et~al.}(2024)\citenamefont{Belvedere,
  Englert, Kogler, and Spannowsky}}]{Belvedere:2024wzg}
\bibinfo{author}{\bibfnamefont{A.}~\bibnamefont{Belvedere}},
  \bibinfo{author}{\bibfnamefont{C.}~\bibnamefont{Englert}},
  \bibinfo{author}{\bibfnamefont{R.}~\bibnamefont{Kogler}}, \bibnamefont{and}
  \bibinfo{author}{\bibfnamefont{M.}~\bibnamefont{Spannowsky}}, ``{Dispelling
  the $\sqrt{L}$ myth for the High-Luminosity LHC},''  (\bibinfo{year}{2024}),
  \eprint{2402.07985}.

\bibitem[{\citenamefont{Butler et~al.}(2023)}]{Butler:2023eah}
\bibinfo{author}{\bibfnamefont{J.~N.} \bibnamefont{Butler}}
  \bibnamefont{et~al.}, ``{Report of the 2021 U.S. Community Study on the
  Future of Particle Physics (Snowmass 2021) Summary Chapter},''
  (\bibinfo{year}{2023}), \eprint{2301.06581}.

\bibitem[{\citenamefont{\text{ATLAS and CMS
  Collaboration}}(2019)}]{ATLAS:2019mfr}
\bibinfo{author}{\bibnamefont{\text{ATLAS and CMS Collaboration}}}, ``{Report
  on the physics at the HL-LHC, and perspectives for the HE-LHC: Collection of
  notes from ATLAS and CMS},'' \bibinfo{journal}{CERN Yellow Rep. Monogr.}
  \textbf{\bibinfo{volume}{7}}, \bibinfo{pages}{Addendum}
  (\bibinfo{year}{2019}), \eprint{1902.10229}.

\bibitem[{\citenamefont{Bambade et~al.}(2019)}]{Bambade:2019fyw}
\bibinfo{author}{\bibfnamefont{P.}~\bibnamefont{Bambade}} \bibnamefont{et~al.},
  ``{The International Linear Collider: A Global Project},''
  (\bibinfo{year}{2019}), \eprint{1903.01629}.

\bibitem[{Aic(2018)}]{Aicheler:2018arh}
``{The Compact Linear Collider (CLIC) - Project Implementation Plan},''
  \textbf{\bibinfo{volume}{4/2018}} (\bibinfo{year}{2018}),
  \eprint{1903.08655}.

\bibitem[{\citenamefont{Bai et~al.}(2021)}]{Bai:2021rdg}
\bibinfo{author}{\bibfnamefont{M.}~\bibnamefont{Bai}} \bibnamefont{et~al.}, in
  \emph{\bibinfo{booktitle}{{Snowmass 2021}}} (\bibinfo{year}{2021}),
  \eprint{2110.15800}.

\bibitem[{\citenamefont{de~Lima and Logan}(2024)}]{deLima:2024hnk}
\bibinfo{author}{\bibfnamefont{C.~H.} \bibnamefont{de~Lima}} \bibnamefont{and}
  \bibinfo{author}{\bibfnamefont{H.~E.} \bibnamefont{Logan}}, ``{Can CP be
  conserved in the two-Higgs-doublet model?},''  (\bibinfo{year}{2024}),
  \eprint{2403.17052}.

\bibitem[{\citenamefont{Alwall et~al.}(2014)\citenamefont{Alwall, Frederix,
  Frixione, Hirschi, Maltoni, Mattelaer, Shao, Stelzer, Torrielli, and
  Zaro}}]{Alwall:2014hca}
\bibinfo{author}{\bibfnamefont{J.}~\bibnamefont{Alwall}},
  \bibinfo{author}{\bibfnamefont{R.}~\bibnamefont{Frederix}},
  \bibinfo{author}{\bibfnamefont{S.}~\bibnamefont{Frixione}},
  \bibinfo{author}{\bibfnamefont{V.}~\bibnamefont{Hirschi}},
  \bibinfo{author}{\bibfnamefont{F.}~\bibnamefont{Maltoni}},
  \bibinfo{author}{\bibfnamefont{O.}~\bibnamefont{Mattelaer}},
  \bibinfo{author}{\bibfnamefont{H.~S.} \bibnamefont{Shao}},
  \bibinfo{author}{\bibfnamefont{T.}~\bibnamefont{Stelzer}},
  \bibinfo{author}{\bibfnamefont{P.}~\bibnamefont{Torrielli}},
  \bibnamefont{and} \bibinfo{author}{\bibfnamefont{M.}~\bibnamefont{Zaro}},
  ``{The automated computation of tree-level and next-to-leading order
  differential cross sections, and their matching to parton shower
  simulations},'' \bibinfo{journal}{JHEP} \textbf{\bibinfo{volume}{07}},
  \bibinfo{pages}{079} (\bibinfo{year}{2014}), \eprint{1405.0301}.

\bibitem[{\citenamefont{Bierlich et~al.}(2022)}]{Bierlich:2022pfr}
\bibinfo{author}{\bibfnamefont{C.}~\bibnamefont{Bierlich}}
  \bibnamefont{et~al.}, ``{A comprehensive guide to the physics and usage of
  PYTHIA 8.3},'' \bibinfo{journal}{SciPost Phys. Codeb.}
  \textbf{\bibinfo{volume}{2022}}, \bibinfo{pages}{8} (\bibinfo{year}{2022}),
  \eprint{2203.11601}.

\bibitem[{\citenamefont{Bierlich et~al.}(2020)}]{Bierlich:2019rhm}
\bibinfo{author}{\bibfnamefont{C.}~\bibnamefont{Bierlich}}
  \bibnamefont{et~al.}, ``{Robust Independent Validation of Experiment and
  Theory: Rivet version 3},'' \bibinfo{journal}{SciPost Phys.}
  \textbf{\bibinfo{volume}{8}}, \bibinfo{pages}{026} (\bibinfo{year}{2020}),
  \eprint{1912.05451}.

\bibitem[{\citenamefont{Sirunyan et~al.}(2021)}]{CMS:2021wlt}
\bibinfo{author}{\bibfnamefont{A.~M.} \bibnamefont{Sirunyan}}
  \bibnamefont{et~al.} (\bibinfo{collaboration}{CMS}), ``{Search for charged
  Higgs bosons produced in vector boson fusion processes and decaying into
  vector boson pairs in proton\textendash{}proton collisions at $\sqrt{s} =
  13\,{\text {TeV}} $},'' \bibinfo{journal}{Eur. Phys. J. C}
  \textbf{\bibinfo{volume}{81}}, \bibinfo{pages}{723} (\bibinfo{year}{2021}),
  \eprint{2104.04762}.

\bibitem[{\citenamefont{Abbiendi et~al.}(2013)}]{ALEPH:2013htx}
\bibinfo{author}{\bibfnamefont{G.}~\bibnamefont{Abbiendi}} \bibnamefont{et~al.}
  (\bibinfo{collaboration}{ALEPH, DELPHI, L3, OPAL, LEP}), ``{Search for
  Charged Higgs bosons: Combined Results Using LEP Data},''
  \bibinfo{journal}{Eur. Phys. J. C} \textbf{\bibinfo{volume}{73}},
  \bibinfo{pages}{2463} (\bibinfo{year}{2013}), \eprint{1301.6065}.

\bibitem[{\citenamefont{Sirunyan et~al.}(2019)}]{CMS:2019bfg}
\bibinfo{author}{\bibfnamefont{A.~M.} \bibnamefont{Sirunyan}}
  \bibnamefont{et~al.} (\bibinfo{collaboration}{CMS}), ``{Search for charged
  Higgs bosons in the H$^{\pm}$ $\to$ $\tau^{\pm}\nu_\tau$ decay channel in
  proton-proton collisions at $\sqrt{s} =$ 13 TeV},'' \bibinfo{journal}{JHEP}
  \textbf{\bibinfo{volume}{07}}, \bibinfo{pages}{142} (\bibinfo{year}{2019}),
  \eprint{1903.04560}.

\bibitem[{\citenamefont{Aaboud et~al.}(2018)}]{ATLAS:2018gfm}
\bibinfo{author}{\bibfnamefont{M.}~\bibnamefont{Aaboud}} \bibnamefont{et~al.}
  (\bibinfo{collaboration}{ATLAS}), ``{Search for charged Higgs bosons decaying
  via $H^{\pm} \to \tau^{\pm}\nu_{\tau}$ in the $\tau$+jets and $\tau$+lepton
  final states with 36 fb$^{-1}$ of $pp$ collision data recorded at $\sqrt{s} =
  13$ TeV with the ATLAS experiment},'' \bibinfo{journal}{JHEP}
  \textbf{\bibinfo{volume}{09}}, \bibinfo{pages}{139} (\bibinfo{year}{2018}),
  \eprint{1807.07915}.

\bibitem[{\citenamefont{Aad et~al.}(2023)}]{ATLAS:2023bzb}
\bibinfo{author}{\bibfnamefont{G.}~\bibnamefont{Aad}} \bibnamefont{et~al.}
  (\bibinfo{collaboration}{ATLAS}), ``{Search for a light charged Higgs boson
  in $t \rightarrow H^{\pm}b$ decays, with $H^{\pm} \rightarrow cb$, in the
  lepton+jets final state in proton-proton collisions at $\sqrt{s}=13$ TeV with
  the ATLAS detector},'' \bibinfo{journal}{JHEP} \textbf{\bibinfo{volume}{09}},
  \bibinfo{pages}{004} (\bibinfo{year}{2023}), \eprint{2302.11739}.

\bibitem[{\citenamefont{Aad et~al.}(2021)}]{ATLAS:2021upq}
\bibinfo{author}{\bibfnamefont{G.}~\bibnamefont{Aad}} \bibnamefont{et~al.}
  (\bibinfo{collaboration}{ATLAS}), ``{Search for charged Higgs bosons decaying
  into a top quark and a bottom quark at $ \sqrt{\mathrm{s}} $ = 13 TeV with
  the ATLAS detector},'' \bibinfo{journal}{JHEP} \textbf{\bibinfo{volume}{06}},
  \bibinfo{pages}{145} (\bibinfo{year}{2021}), \eprint{2102.10076}.

\bibitem[{\citenamefont{Sirunyan et~al.}(2020)}]{CMS:2019rlz}
\bibinfo{author}{\bibfnamefont{A.~M.} \bibnamefont{Sirunyan}}
  \bibnamefont{et~al.} (\bibinfo{collaboration}{CMS}), ``{Search for a charged
  Higgs boson decaying into top and bottom quarks in events with electrons or
  muons in proton-proton collisions at $ \sqrt{\mathrm{s}} $ = 13 TeV},''
  \bibinfo{journal}{JHEP} \textbf{\bibinfo{volume}{01}}, \bibinfo{pages}{096}
  (\bibinfo{year}{2020}), \eprint{1908.09206}.

\bibitem[{\citenamefont{Tumasyan et~al.}(2022{\natexlab{b}})}]{CMS:2022nty}
\bibinfo{author}{\bibfnamefont{A.}~\bibnamefont{Tumasyan}} \bibnamefont{et~al.}
  (\bibinfo{collaboration}{CMS}), ``{Inclusive nonresonant multilepton probes
  of new phenomena at $\sqrt s$=13\,\,TeV},'' \bibinfo{journal}{Phys. Rev. D}
  \textbf{\bibinfo{volume}{105}}, \bibinfo{pages}{112007}
  (\bibinfo{year}{2022}{\natexlab{b}}), \eprint{2202.08676}.

\bibitem[{\citenamefont{Ismail et~al.}(2021)\citenamefont{Ismail, Keeshan,
  Logan, and Wu}}]{Ismail:2020kqz}
\bibinfo{author}{\bibfnamefont{A.}~\bibnamefont{Ismail}},
  \bibinfo{author}{\bibfnamefont{B.}~\bibnamefont{Keeshan}},
  \bibinfo{author}{\bibfnamefont{H.~E.} \bibnamefont{Logan}}, \bibnamefont{and}
  \bibinfo{author}{\bibfnamefont{Y.}~\bibnamefont{Wu}}, ``{Benchmark for LHC
  searches for low-mass custodial fiveplet scalars in the Georgi-Machacek
  model},'' \bibinfo{journal}{Phys. Rev. D} \textbf{\bibinfo{volume}{103}},
  \bibinfo{pages}{095010} (\bibinfo{year}{2021}), \eprint{2003.05536}.

\bibitem[{\citenamefont{Degrande et~al.}(2017)\citenamefont{Degrande, Hartling,
  and Logan}}]{Degrande:2017naf}
\bibinfo{author}{\bibfnamefont{C.}~\bibnamefont{Degrande}},
  \bibinfo{author}{\bibfnamefont{K.}~\bibnamefont{Hartling}}, \bibnamefont{and}
  \bibinfo{author}{\bibfnamefont{H.~E.} \bibnamefont{Logan}}, ``{Scalar decays
  to $\gamma\gamma$, $Z\gamma$, and $W\gamma$ in the Georgi-Machacek model},''
  \bibinfo{journal}{Phys. Rev. D} \textbf{\bibinfo{volume}{96}},
  \bibinfo{pages}{075013} (\bibinfo{year}{2017}), \bibinfo{note}{[Erratum:
  Phys.Rev.D 98, 019901 (2018)]}, \eprint{1708.08753}.

\bibitem[{\citenamefont{Ashanujjaman et~al.}(2023)\citenamefont{Ashanujjaman,
  Ghosh, and Sahu}}]{PhysRevD.107.015018}
\bibinfo{author}{\bibfnamefont{S.}~\bibnamefont{Ashanujjaman}},
  \bibinfo{author}{\bibfnamefont{K.}~\bibnamefont{Ghosh}}, \bibnamefont{and}
  \bibinfo{author}{\bibfnamefont{R.}~\bibnamefont{Sahu}}, ``Low-mass doubly
  charged higgs bosons at the lhc,'' \bibinfo{journal}{Phys. Rev. D}
  \textbf{\bibinfo{volume}{107}}, \bibinfo{pages}{015018}
  (\bibinfo{year}{2023}),
  \urlprefix\url{https://link.aps.org/doi/10.1103/PhysRevD.107.015018}.

\end{thebibliography}

\end{document}